\documentclass[twocolumn,appendixfloats,tighten]{aastex631}
\usepackage[T1]{fontenc}
\usepackage{ae,aecompl}

\usepackage{amsmath}	
\usepackage{amssymb}	

\usepackage{ctable}
\usepackage{url}
\usepackage{xspace}
\usepackage[normalem]{ulem}
\usepackage{hyperref}
\hypersetup{linkcolor=red,citecolor=blue,filecolor=cyan,urlcolor=magenta}
\usepackage[all]{hypcap}
\usepackage{graphicx}
\usepackage{comment}

\usepackage{amssymb}
\usepackage{pifont}

\usepackage{etoolbox}
\newtoggle{ApJFigs}
\togglefalse{ApJFigs}

\def\msun{{\rm\,M_\odot}}

\def\msun{{\rm\,M_\odot}}

\newcommand{\kms}{\, {\rm km\, s}^{-1}}
\newcommand{\gconst}{\mathcal{G}}

\newcommand{\be}{\begin{equation}}
\newcommand{\ee}{\end{equation}}

\newcommand{\au}{\mathrm{au}}

\newcommand{\pc}{\mathrm{pc}}
\def\h2{${\rm\,H_2}$}

\newcommand{\F}{Fig.}
\newcommand{\Fs}{Figs.}
\newcommand{\eq}{equation}
\newcommand{\Eq}{Equation}

\newcommand{\myr}{\mathrm{Myr}}
\newcommand{\yr}{\mathrm{yr}}
\newcommand{\gyr}{\mathrm{Gyr}}

\newcommand{\mse}{\textsc{MSE}}
\newcommand{\pk}{PK12}

\usepackage{color}

\begin{document}

\title{Return of the TEDI: revisiting the Triple Evolution Dynamical Instability channel in triple stars} 

\shortauthors{Hamers et al.}

\author[0000-0003-1004-5635]{Adrian S. Hamers}
\affiliation{Max-Planck-Institut f\"{u}r Astrophysik, Karl-Schwarzschild-Str. 1, 85741 Garching, Germany}

\author[0000-0002-5004-199X]{Hagai B. Perets}
\affiliation{Faculty of Physics, Technion -- Israel Institute of Technology, Haifa, 3200003, Israel}

\author[0000-0003-2377-9574]{Todd A. Thompson}
\affiliation{Department of Astronomy and the Center for Cosmology and Astro-Particle Physics, Ohio State University, 140 West 18th Avenue, Columbus, OH 43210, USA}

\author[0000-0001-5853-6017]{Patrick Neunteufel}
\affiliation{Max-Planck-Institut f\"{u}r Astrophysik, Karl-Schwarzschild-Str. 1, 85741 Garching, Germany}

\begin{abstract} 
Triple-star systems exhibit a phenomenon known as the Triple Evolution Dynamical Instability (TEDI), in which mass loss in evolving triples triggers short-term dynamical instabilities, potentially leading to collisions of stars, exchanges, and ejections. Previous work has shown that the TEDI is an important pathway to head-on stellar collisions in the Galaxy, significantly exceeding the rate of collisions due to random encounters in globular clusters. Here, we revisit the TEDI evolutionary pathway using state-of-the-art population synthesis methods that self-consistently take into account stellar evolution and binary interactions, as well as gravitational dynamics and perturbations from passing stars in the field. We find Galactic TEDI-induced collision rates on the order of $10^{-4} \,\yr^{-1}$, consistent with previous studies which were based on more simplified methods. The majority of TEDI-induced collisions involve main sequence stars, potentially producing blue straggler stars. Collisions are also possible involving more evolved stars, potentially producing eccentric post-common-envelope systems, and white dwarfs collisions leading to Type Ia supernovae (although the latter with a negligible contribution to the Galactic rate). In our simulations, the TEDI is not only triggered by adiabatic wind mass loss, but also by Roche lobe overflow in the inner binary: when the donor star becomes less massive than the accretor, the inner binary orbit widens, triggering triple dynamical instability. We find that collision rates are increased by $\sim 17\%$ when fly-bys in the field are taken into account. In addition to collisions, we find that the TEDI produces $\sim 10^{-4}\,\yr^{-1}$ of unbound stars, although none with escape speeds in excess of $10^3\,\kms$.
\end{abstract}

\section{Introduction}
\label{sect:intro}
Hierarchical triple systems consist of two bodies in an inner orbit, with a third object orbiting the inner orbit in a wider orbit. They are well known for von Zeipel-Lidov-Kozai (ZLK) oscillations, which arise from long-term gravitational exchanges of angular momentum between the inner and outer orbits (\citealt{1910AN....183..345V,1962P&SS....9..719L,1962AJ.....67..591K}; see \citealt{2016ARA&A..54..441N,2017ASSL..441.....S,2019MEEP....7....1I} for reviews). ZLK oscillations can give rise to high eccentricities in the inner orbit, potentially enhancing the strength of tidal interactions, triggering mass transfer, or even causing physical collisions. These effects have important implications in a wide range of astrophysical contexts, including (in combination with tidal effects) the production of short-period binaries (e.g., \citealt{1979A&A....77..145M,1998MNRAS.300..292K,2001ApJ...562.1012E,2006Ap&SS.304...75E,2007ApJ...669.1298F,2014ApJ...793..137N}), and hot Jupiters (e.g., \citealt{2003ApJ...589..605W,2007ApJ...669.1298F,2012ApJ...754L..36N,2015ApJ...799...27P,2016MNRAS.456.3671A,2016ApJ...829..132P,2017ApJ...835L..24H}). Other implications include enhancing mergers of compact objects leading to gravitational wave (GW) events (e.g., \citealt{2002ApJ...578..775B,2011ApJ...741...82T,2013MNRAS.430.2262H,2014MNRAS.439.1079A,2017ApJ...841...77A,2017ApJ...836...39S,2017ApJ...846L..11L,2018ApJ...863...68L,2018ApJ...865....2H,2018ApJ...856..140H,2018ApJ...853...93R,2018ApJ...864..134R,2018A&A...610A..22T,2019MNRAS.486.4443F}), affecting the evolution of protoplanetary or accretion disks in binaries (e.g., \citealt{2014ApJ...792L..33M,2015ApJ...813..105F,2017MNRAS.467.1957Z,2017MNRAS.469.4292L,2018MNRAS.477.5207Z,2019MNRAS.485..315F,2019MNRAS.489.1797M}), triggering white dwarf pollution by planets (e.g., \citealt{2016MNRAS.462L..84H,2017ApJ...834..116P}), and producing blue straggler stars (e.g., \citealt{2009ApJ...697.1048P,2016ApJ...816...65A,2016MNRAS.460.3494S,2019MNRAS.488..728F}). 

However, in {\it stellar} triples, strong interactions such as collisions not only arise because of purely gravitational dynamical evolution, but they can also occur during phases of dynamical instability triggered by changes in the orbits as a result of stellar or binary evolution-induced mass loss or mass transfer \citep{1994MNRAS.270..936K,1999ApJ...511..324I,2010arXiv1001.0581P,2011MNRAS.412.2763F,2011ApJ...734...55P,2012ApJ...760...99P}. The pioneering work of \citeauthor{2012ApJ...760...99P} (\citeyear{2012ApJ...760...99P}; hereafter \pk) showed quantitatively how, in stellar triples, mass loss from stellar winds can lead to a system becoming unstable, ultimately causing head-on collisions between stars. 

In particular, if a star in a triple is losing mass, this will affect one or both orbits. Under the assumption of adiabatic mass loss (i.e., the wind mass loss time-scale is much longer than the orbital period), both $a_1(m_1+m_2)$ and $a_2(m_1+m_2+m_3)$ will remain constant \citep{1963Icar....2..440H}, where $a_1$ and $a_2$ are the inner and outer orbital semimajor axes, respectively, and $m_1$, $m_2$, and $m_3$ are the inner binary primary and secondary, and tertiary masses, respectively\footnote{The eccentricities remain constant under the assumption of adiabatic mass loss.}. If one of the stars in the inner binary is losing mass, this implies that the ratio $a_2/a_1$ {\it decreases}, i.e., the triple becomes more compact in a hierarchical sense. This can trigger stronger secular evolution (i.e., higher inner orbital eccentricities; e.g., \citealt{2013ApJ...766...64S,2014ApJ...794..122M}), but can also destabilise the system. A dynamical instability phase is typically chaotic; the hierarchy of the system could be broken, and stars (not only the inner binary stars) could collide, or be ejected from the system. \pk~showed that this Triple Evolution Dynamical Instability (TEDI) pathway leads to a comparatively high rate of stellar head-on collisions in the Galaxy, namely on the order of $10^{-4} \, \yr^{-1}$, which is $\sim 30$ higher than the rate expected from random encounters in Galactic globular clusters. For reference, the inferred rate of all stellar mergers in the Galaxy (including the initiation of common envelope phases) is much higher at $\sim 0.5 \, \yr^{-1}$ \citep{2014MNRAS.443.1319K}, but the TEDI is unique in its ability to produce head-on collisions in the field, with potentially unique signatures.

Stellar mergers are interesting for a variety of reasons. Merging main-sequence (MS) stars can produce blue straggler stars, which are MS stars that appear to be too blue and luminous compared to the MS turnoff point of the cluster that they reside in \citep[e.g.,][]{1953AJ.....58...61S,1993PASP..105.1081S,1995ARA&A..33..133B,1997ApJ...487..290S,1999ApJ...513..428S,2001ApJ...548..323S,2002MNRAS.332...49S,2013ApJ...777..106C}. Also, in clusters, MS mergers can lead to very massive stars that could become intermediate-mass BHs \citep[e.g.,][]{1999A&A...348..117P,2002ASPC..267..193B,2010MNRAS.402..105G,2013MNRAS.430.1018F,2016MNRAS.459.3432M,2020ApJ...892...36T,2021MNRAS.501.5257R}. Collisions involving more evolved stars such as giant stars can lead to common-envelope-like evolution in which the envelope is removed, and a tight and possibly eccentric binary system remains, especially if the collision is head on or highly eccentric \citep[e.g.,][]{1999MNRAS.308..257B,2008ApJ...678..922Y}. 

\pk~studied the TEDI by modelling the evolution of an isolated binary-star system with the population synthesis codes \textsc{BSE} \citep{2002MNRAS.329..897H}, as well as taking into account single stellar evolution of the tertiary star with \textsc{SSE}  \citep{2000MNRAS.315..543H}. They assumed that the inner binary evolution is completely decoupled from the tertiary star. In post processing, they modelled the evolution of the outer orbit semimajor axis assuming adiabatic mass loss and using the recorded history of the masses of all three stars, as well as the inner orbital semimajor axis (as dictated by \textsc{BSE}, and in isolation). From the resulting time series of the masses and semimajor axes, it was determined which systems would become dynamically unstable, and the subsequent short-term dynamical evolution was modelled with a direct $N$-body integrator, keeping track of stellar collisions. 

Although highly innovative, \pk~made a number of key approximations. Most importantly, they assumed that the inner binary evolution is decoupled from the tertiary star until dynamical instability occurs. This assumption can break down, in particular if the inner binary is excited to high eccentricity due to ZLK oscillations. For example, a TEDI system identified by \pk~could potentially already undergo high-eccentricity ZLK oscillations before the orbits start expanding because of mass loss, driving a stellar merger in the inner orbit and avoiding a later instability phase \citep[e.g.,][]{2013ApJ...766...64S,2014ApJ...794..122M}. Other consequences of ZLK oscillations include driving strong tidal orbital shrinkage and common-envelope (CE) evolution in the inner binary, after which ZLK oscillations are likely quenched \citep[e.g.,][]{2013MNRAS.430.2262H}. 

Furthermore, \pk~did not include the effects of stars passing by the triple system. Such fly-bys can be important for wide orbits in the field. In particular, for wide binary systems, they can drive stellar collisions by making the binary orbit highly eccentric \citep[e.g.,][]{2014ApJ...782...60K,2019ApJ...887L..36M}. It is unclear how these fly-bys affect the TEDI. 

Also, \pk~focussed on stellar collisions, whereas other outcomes of the TEDI include ejections of stars from the triple system, and exchange interactions. Such scattering interactions can be triggered by the TEDI even in isolated triples, whereas stable hierarchical triples otherwise require external incoming star(s) for strong scattering to occur, e.g., in dense stellar systems \citep[e.g.,][]{2016MNRAS.456.4219A,2016ApJ...816...65A,2020ApJ...900...16F,2020ApJ...903...67M}.

Here, we present a more self-consistent study of the TEDI scenario in which we simultaneously take into account stellar evolution, binary interactions such as tidal evolution, mass transfer and CE evolution, and gravitational dynamics (including the effects of fly-bys, and following the outcome of instability with direct $N$-body integration). This is achieved with the new Multiple Stellar Evolution (\mse) code \citep{2021MNRAS.502.4479H}. Our focus is on both `clean', head-on collisions induced by the TEDI (with the instantaneous separation being less than the sum of the radii and involving stars without extended envelopes such as MS stars and compact objects), and CE evolution triggered by the TEDI (this can occur when at least one of two physically-colliding stars has an extended envelope). In the latter case, a tight and potentially eccentric binary could remain (with a more distant tertiary star).

The structure of this paper is as follows. In \S\ref{sect:meth}, we briefly describe the population synthesis code used in this work. We give a number of examples of different varieties of TEDI evolution in \S\ref{sect:ex}. In \S\ref{sect:popmeth}, we describe the assumptions made to generate an initial population of triples, and in \S\ref{sect:results}, we show the results from the population synthesis calculations. We discuss our findings in \S\ref{sect:dis}, and conclude in \S\ref{sect:conclusions}.

\section{Methodology}
\label{sect:meth}
Here, we give a brief overview of the population synthesis code used in this work, \mse\footnote{The version used for the population synthesis simulations in this work is v0.83.}. For more details on \mse, we refer to \citet{2021MNRAS.502.4479H}.

\subsection{Dynamics}
\label{sect:meth:dyn}
The \mse~code models the evolution of an arbitrary number of stars in an initial hierarchical configuration. Gravitational dynamical evolution is taken into account via two methods: (1) secular (orbit-averaged) integration for sufficiently hierarchical systems, based on the formalism of \citet{2016MNRAS.459.2827H,2018MNRAS.476.4139H,2020MNRAS.494.5492H}, and (2) direct $N$-body integration for cases in which the secular approach breaks down using the algorithmic chain regularization code \textsc{MSTAR} \citep{2020MNRAS.492.4131R}. 

The secular integrations include tidal evolution under the assumption of equilibrium tides \citep{1981A&A....99..126H,1998ApJ...499..853E}, where the efficiency of tidal dissipation (for damping in stars dominated by convective, radiative, or degenerate regions) is computed using the prescription of \citet{2002MNRAS.329..897H}. Furthermore, the evolution of the stellar spins is included taking into account the relevant terms for tidal bulges and rotation from \citet{1998ApJ...499..853E}; spin-orbit misalignment is allowed (by default, stars are assumed to form with zero spin-orbit misalignment). Also included are post-Newtonian (PN) terms to the 1PN and 2.5PN orders in all orbits, neglecting PN `interaction' terms that can arise between different orbits (e.g., \citealt{2013ApJ...773..187N,2014CQGra..31x4001W,2020PhRvD.102f4033L}). The lowest-order PN spin-orbit coupling terms (that describe precession of the spins around the orbit, and are mostly important for compact object systems) are also included. 

The direct $N$-body integrations currently do not include additional acceleration terms that describe tidal evolution. However, collision detection is implemented. Here, we consider stellar collisions when the instantaneous relative separation between two stars is less than the sum of their radii. For compact objects (WDs, NSs, and BHs), we adopt another condition for `collisions', namely the instantaneous relative separation should be less than 100 times the mutual gravitational radius, $r_\mathrm{g}=\gconst (m_1+m_2)/c^2$, where $m_1$ and $m_2$ are the masses of the two compact objects, and $\gconst$ and $c$ are the gravitational constant and speed of light, respectively. The latter condition for compact objects is included for computational reasons, since the higher-order PN terms become computationally very prohibitive at small separations, and the inspiral time due to GW emission at such small separations is very short (see, e.g., \Eq~108 in \citealt{2021MNRAS.502.4479H}). The \textsc{MSTAR} code includes PN terms to 1PN, 2PN, 2.5PN, 3PN, and 3.5PN orders (e.g., \citealt{2004PhRvD..69j4021M,2006LRR.....9....3W}), as well as spin-orbit, spin-spin, and quadrupole terms \citep{1975PhRvD..12..329B,1995PhRvD..52..821K}. 

Often, the system is dynamically stable, and the secular approximation applies. However, evolutionary processes such as mass loss from stellar evolution can destabilise a system, or could void the validity of the secular approximation. In \mse, it is continuously checked whether the system is still stable using the stability criterion of \citet{2001MNRAS.321..398M}, i.e.,
\begin{align}
\label{eq:dynstab}
\nonumber \frac{a_\mathrm{out}(1-e_\mathrm{out})}{a_\mathrm{in}} &> 2.8 \, \left [ (1+q_\mathrm{out}) \frac{1+e_\mathrm{out}}{\sqrt{1-e_\mathrm{out}}} \right ]^{2/5} \\
&\quad \times \left (1-0.3\, \frac{\Phi}{\pi} \right ).
\end{align}
This criterion is assumed to apply to any pair of orbits in the system; the subscripts `in' and `out' refer to the inner and outer orbits for such a pair. The mass ratio $q_\mathrm{out}$ is defined as $q_\mathrm{out}\equiv (M_\mathrm{out}-M_\mathrm{in})/M_\mathrm{in}$, where $M_\mathrm{in}$ is the mass of all bodies contained within the inner orbit, and $M_\mathrm{out}$ is the mass of all bodies contained within the outer orbit (including those in the inner orbit). The angle $\Phi$ is the mutual inclination between the pair of orbits (expressed in radians). 

The secular approximation is likely to break down if \eq~(\ref{eq:dynstab}) is satisfied. However, other, less stringent conditions exist in which the secular approximation breaks down, namely when the secular time-scale becomes comparable to one of the orbital periods in the system \citep[e.g.,][]{2012ApJ...757...27A,2014ApJ...781...45A,2016MNRAS.458.3060L,2018MNRAS.481.4907G,2018MNRAS.481.4602L,2019MNRAS.490.4756L,2020MNRAS.494.5492H}. The \mse~code checks for this `semisecular' regime by comparing the instantaneous time-scale for secular evolution to change the specific angular momentum of any orbit $k$, $\jmath_k$, to the orbital period $P_{\mathrm{orb},\,k}$, i.e.,
\begin{align}
\label{eq:semisec}
t_{\jmath_k} \equiv \left | \frac{1}{\jmath_k} \left ( \frac{\mathrm{d} \jmath_k}{\mathrm{d} t} \right )_\mathrm{sec} \right |^{-1} < P_{\mathrm{orb},\,k}.
\end{align}

If either an instability occurs according to \eq~(\ref{eq:dynstab}) or the semisecular regime is entered, \mse~switches to direct $N$-body integration. Subsequently, the $N$-body system is analysed and if (and only if) a stable hierarchical (sub)configuration is identified for the entire system (this can include unbound objects), it will switch back to secular integration.

\subsection{Stellar evolution}
\label{sect:meth:stel}
Stellar evolution in \mse~is modelled using fast analytic fitting formulae to detailed stellar evolution models from \citet{2000MNRAS.315..543H}. These tracks include information on large-scale parameters such as total mass, radius, luminosity, and global properties of the core (if present). Also included is mass loss due to stellar winds, and spin-down due to magnetic braking \citep{1983ApJ...275..713R}. In our simulations, we adopt Solar metallicity for all stars, $Z_i = 0.02$. In a stellar population, extremely low-mass stars (near the H burning limit of $\approx 0.08\,\msun$) may have lower metallicity, implying different radii. However, the cross section for collisions is dominated by the more massive stars, so we do not expect this caveat to affect our overall results. Simulations with different metallicities are beyond the scope of this paper.

Mass loss due to stellar winds is assumed to affect the orbits adiabatically, i.e., with $m_\mathrm{enc} a_i$ and $e_i$ constant, where $m_\mathrm{enc}$ is the enclosed mass, and $a_i$ and $e_i$ are the orbital semimajor axis and eccentricity, respectively. 

When stars evolve to become NSs or BHs, we assume that mass is lost instantaneously (i.e., the opposite from the adiabatic regime), and take into account the effect of the mass loss on all orbits in the system, assuming no interaction with the lost mass. We also take into account natal kicks for NSs and BHs, by adopting the default `kick distribution model 1' from \citet{2021MNRAS.502.4479H}, in which the natal kick speeds are drawn from a Maxwellian distribution with dispersion $\sigma_{\mathrm{kick}} = 265\,\kms$ for NSs, \citealt{2005MNRAS.360..974H}) and $\sigma_{\mathrm{kick}} = 50\,\kms$ for BHs. We remark that the majority of systems in our population synthesis runs have primary stars with masses less than $8\,\msun$ (cf. \S\ref{sect:popmeth}) that do not become NSs or BHs (at least in isolation), so we do not expect our specific choice of $\sigma_{\mathrm{kick}}$ to significantly affect our results, except for the escape speed distribution (cf. \S\ref{sect:results:vesc}). A detailed study of the more massive range is left for future work. 

\begin{figure*}
\iftoggle{ApJFigs}{
\includegraphics[width=1.1\linewidth,trim = 30mm 30mm 0mm 0mm]{system_17765_mobile}
}{
\includegraphics[width=1.1\linewidth,trim = 30mm 30mm 0mm 0mm]{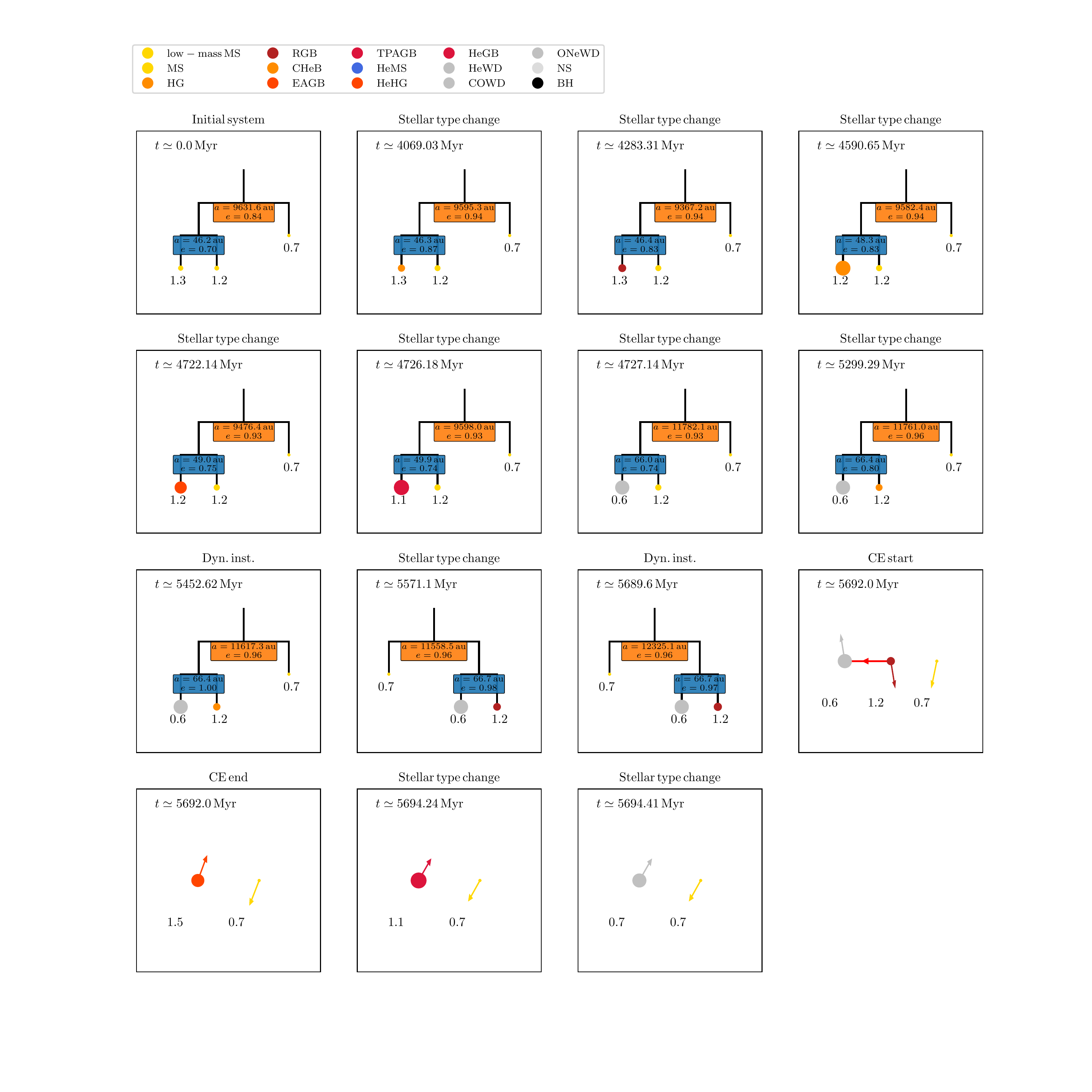}
}
\caption{Mobile diagrams \citep{1968QJRAS...9..388E} at key points in the evolution of an example system of the `traditional' TEDI (cf. \S\ref{sect:ex:1}). The title of each panel describes the event, where dynamical instability is abbreviated as `Dyn. inst.'. Time is indicated in the top left of each panel, and the semimajor axes $a$ and eccentricities $e$ of the orbits are indicated. Colors of the stars correspond to the stellar types (refer to the legend). The arrows emanating from stars (where applicable) indicate the velocity directions, i.e., the unit velocity vectors projected onto the $(x,y)$-plane. The red arrows connecting stars indicate interactions such as RLOF, CE evolution and `clean collisions'.}
\label{fig:ex1_mob}
\end{figure*}

\begin{figure}
\iftoggle{ApJFigs}{
\includegraphics[width=\linewidth,trim = 0mm 0mm 0mm 0mm]{system_17765}
}{
\includegraphics[width=1.1\linewidth,trim = 10mm 0mm 0mm 0mm]{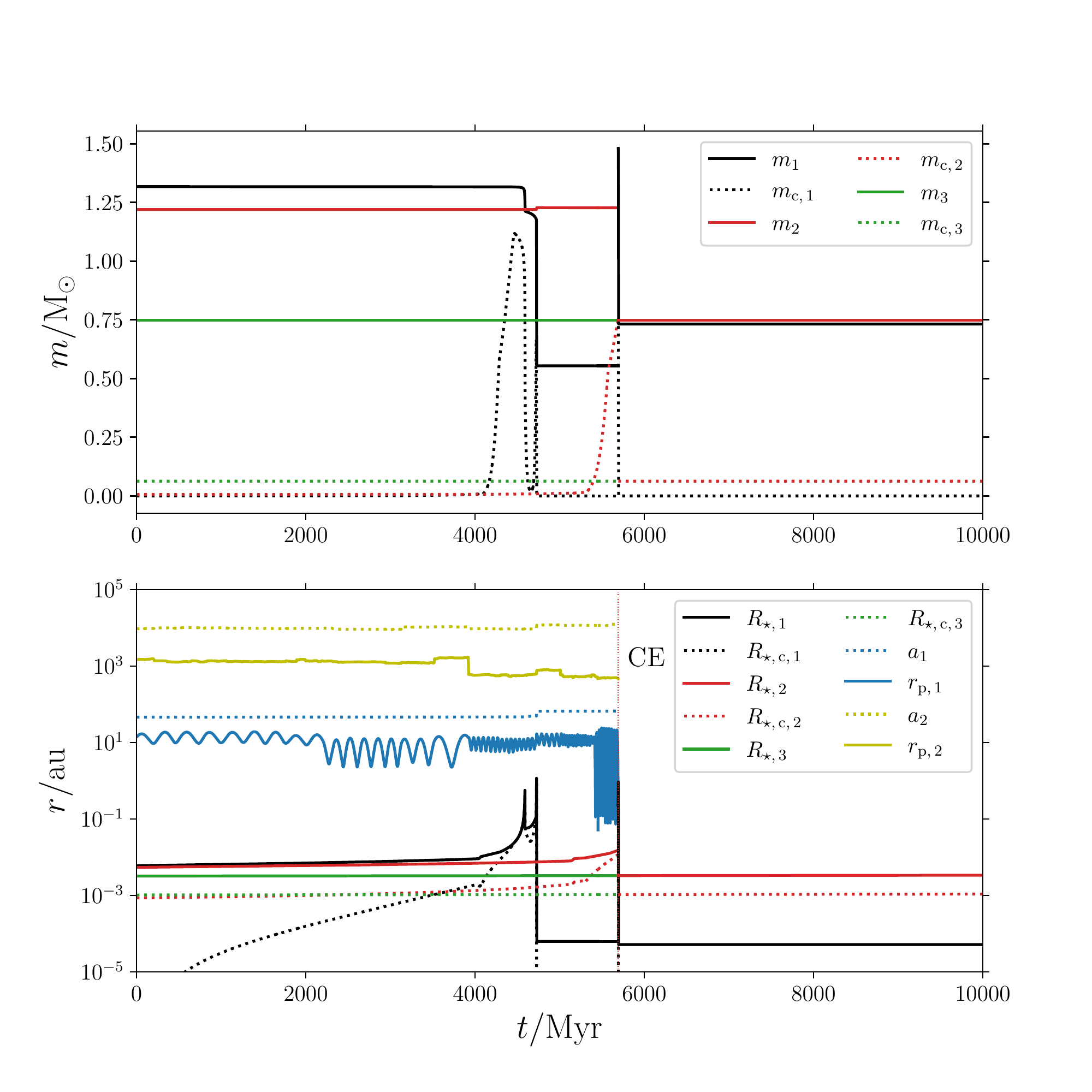}
}
\caption{Masses (top panel), and orbital separations and stellar radii (bottom panel) as a function of time for the example system discussed in \S\ref{sect:ex:1}. In the top panel, black, red, and green lines correspond to the primary, secondary, and tertiary star, respectively. Solid lines show the total stellar masses, and dotted lines the convective core masses. In the bottom panel, the inner and outer orbital semimajor axes (periapsis distances $r_{\mathrm{p},\,i}$) are shown with the upper dotted (solid) lines, where blue (yellow) corresponds to the inner (outer) orbit. The bottom lines show the stellar radii, with the same colors as in the top panel. }
\label{fig:ex1}
\end{figure}

\subsection{Binary interactions}
\label{sect:meth:bin}
The \mse~code includes a number of binary interactions. We check for the condition of Roche lobe overflow (RLOF) for all stars onto companions. If the companion is a single star, `binary' mass transfer applies; otherwise, `triple' mass transfer or triple CE could occur (see \S\ref{sect:meth:tr}).

In the binary case, details of the mass transfer process such as the mass transfer rate, aging/rejuvenation, and the conditions for unstable CE evolution are modelled using similar prescriptions as those used by \citet{2002MNRAS.329..897H}. However, an important difference from \citet{2002MNRAS.329..897H} is the orbital response to mass transfer: \citet{2002MNRAS.329..897H} assumed that tides are always efficient enough to circularise the orbit at the onset of mass transfer. This assumption can break down in triple or higher-order systems, in which eccentricity can be excited secularly \citep[e.g.,][]{2020A&A...640A..16T}. \mse~includes the analytic model of \citet{2019ApJ...872..119H} to describe the orbital response to mass transfer in eccentric orbits. 

Unstable mass transfer can lead to CE evolution; this is taken into account in \mse~by adopting the $\alpha_\mathrm{CE}$-$\lambda$ prescription, similar to \citet{2002MNRAS.329..897H}. Here, we assume $\alpha_\mathrm{CE}=1$ (the default value in \mse). Prescriptions for the outcomes of CE evolution are also adopted from \citet{2002MNRAS.329..897H}. 

In close orbits, accretion of material from stellar winds onto companions can be important. \mse~includes this process of wind accretion by adopting the Bondi-Hoyle-Lyttleton formalism \citep{1939PCPS...35..405H,1944MNRAS.104..273B}. 

\mse~ checks for the onset of direct physical collisions (e.g., as a result of high eccentricity from ZLK oscillations, or after dynamical instability). The properties of the remnant (its new stellar type, mass, etc.) are determined using similar assumptions as those in \citet{2002MNRAS.329..897H}. In particular, in the case when there is one remnant, the new position of the remnant is given by the center of mass position of the two colliding objects, and its new velocity is determined by considering linear momentum conservation. When two MS merge, the resulting MS star is assumed to form without mass loss. In the case of CE evolution, the envelope of the giant star is removed, the core masses of the two objects (one of which can be zero) are added, and the remnant mass (in case of coalescence to a single star) is determined by considering the binding energies of the envelopes prior to and after the CE, and assuming that the remnant radius $R_\star \propto m^{-x}$, where $x$ is given by \eq~(47) of \citet{2000MNRAS.315..543H}.

\subsection{Triple interactions}
\label{sect:meth:tr}
In some systems, an outer star can fill its Roche lobe around an inner companion consisting of two stars (in contrast to `binary' mass transfer). This type of evolution, which can result in stable transfer onto the companion binary, or unstable `triple CE' evolution, is still poorly understood. \mse~models this phase by adopting a number of simplified prescriptions, motivated by more detailed simulations \citep{2014MNRAS.438.1909D,2020MNRAS.498.2957C,2021MNRAS.500.1921G}. 

One of the outcomes of triple CE is a short-term dynamical instability if the core of the donor star ventures too close to the companion binary star, possibly leading to collisions, or ejections of objects \citep{2020MNRAS.498.2957C,2021MNRAS.500.1921G}. This type of evolution could also be considered to be part of the TEDI and, fundamentally, it is taken into account in the prescription in \mse~(see \citealt{2021MNRAS.502.4479H}). In this paper, however, our focus is on collisions following the more `traditional' channel in which instabilities do not involve triple CE evolution. A comprehensive study of the latter phenomenon is beyond the scope of this work and carried out elsewhere \citep{2021arXiv211000024H}.

\subsection{Fly-bys}
\label{sect:meth:fl}
The effects of passing stars in the field (i.e., low-density environments) are taken into account in \mse~by sampling interloping stars during the simulation with a Monte Carlo approach. Based on the stellar density and relative velocity dispersion, perturbers are sampled that impinge on an `encounter sphere' with a large radius $R_\mathrm{enc}$. The effects of the perturber on the multiple system are then computed for impulsive encounters (the latter are most important in the field, although secular encounters dominate in dense stellar systems such as globular clusters, see, e.g., \citealt{1996MNRAS.282.1064H,2019MNRAS.487.5630H}). 

We take into account only encounters with single stars, and not with higher-order multiple systems. Encounters with binaries and higher-order systems can generally be more efficient than those with single objects \citep[e.g.,][]{2015MNRAS.448..344L,2020MNRAS.494..850H}. However, the majority of encounters in the field are wide and in the impulsive limit; in the latter case, we expect additional orbital motion of components within the perturbing system to be small. A quantitative study of the impact of extended perturbers is beyond the scope of this paper.

\section{TEDI examples}
\label{sect:ex}
In this section, we present two examples of the TEDI. These examples were selected from the population synthesis calculations (cf. \S\ref{sect:popmeth}, and  \S\ref{sect:results}; fly-bys were included); their initial conditions are given in Table~\ref{table:exICs}.

\begin{figure*}
\iftoggle{ApJFigs}{
\includegraphics[width=1.1\linewidth,trim = 40mm 100mm 0mm 0mm]{system_26733_mobile}
}{
\includegraphics[width=1.1\linewidth,trim = 40mm 100mm 0mm 0mm]{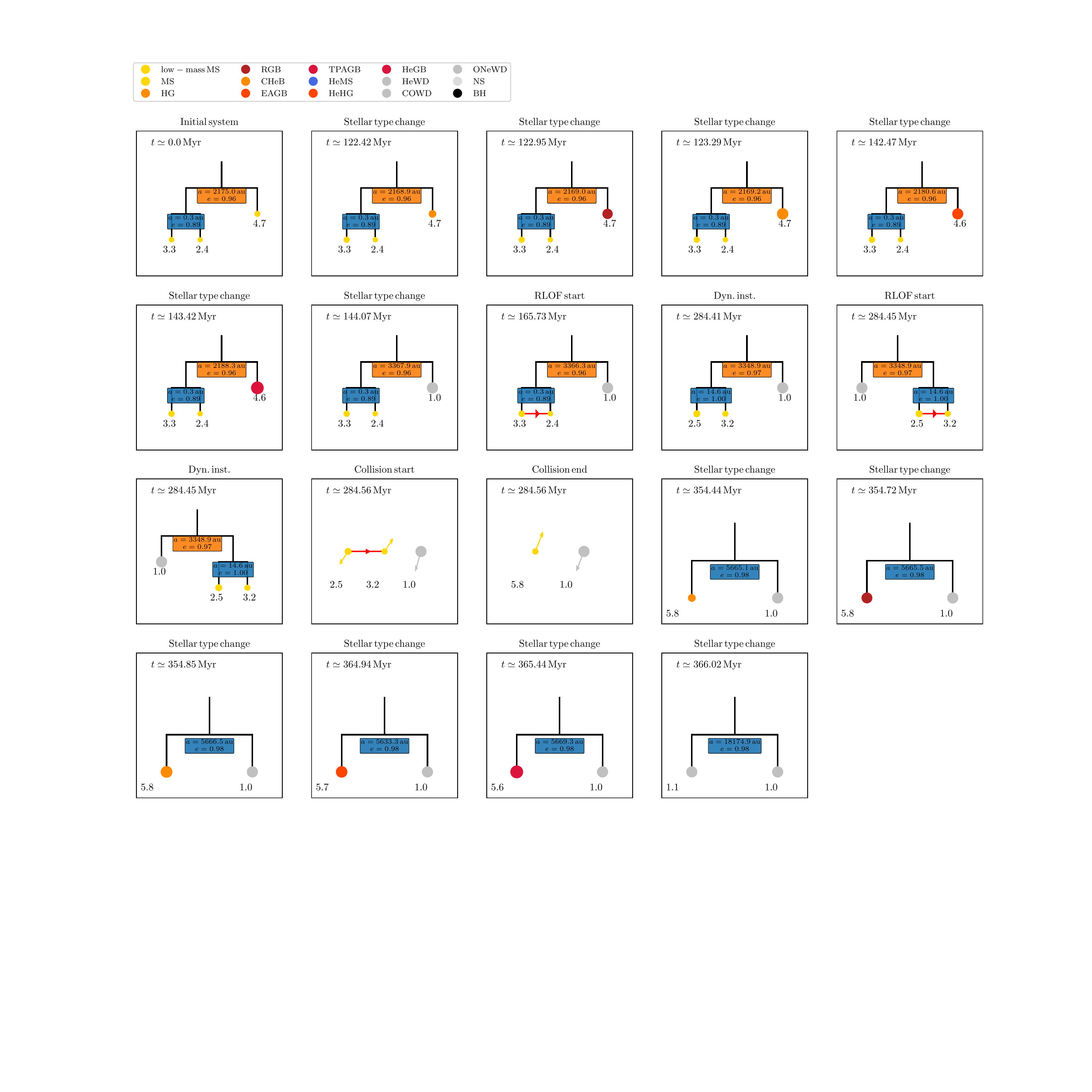}
}
\caption{Mobile diagrams for a system in which TEDI is triggered by RLOF in the inner binary system after the donor becomes less massive than the accretor (cf. \S\ref{sect:ex:2}). In the case of RLOF, the direction of mass transfer is indicated with red arrows. }
\label{fig:ex2_mob}
\end{figure*}

\begin{figure}
\iftoggle{ApJFigs}{
\includegraphics[width=1.1\linewidth,trim = 10mm 0mm 0mm 0mm]{system_26733}
}{
\includegraphics[width=1.1\linewidth,trim = 10mm 0mm 0mm 0mm]{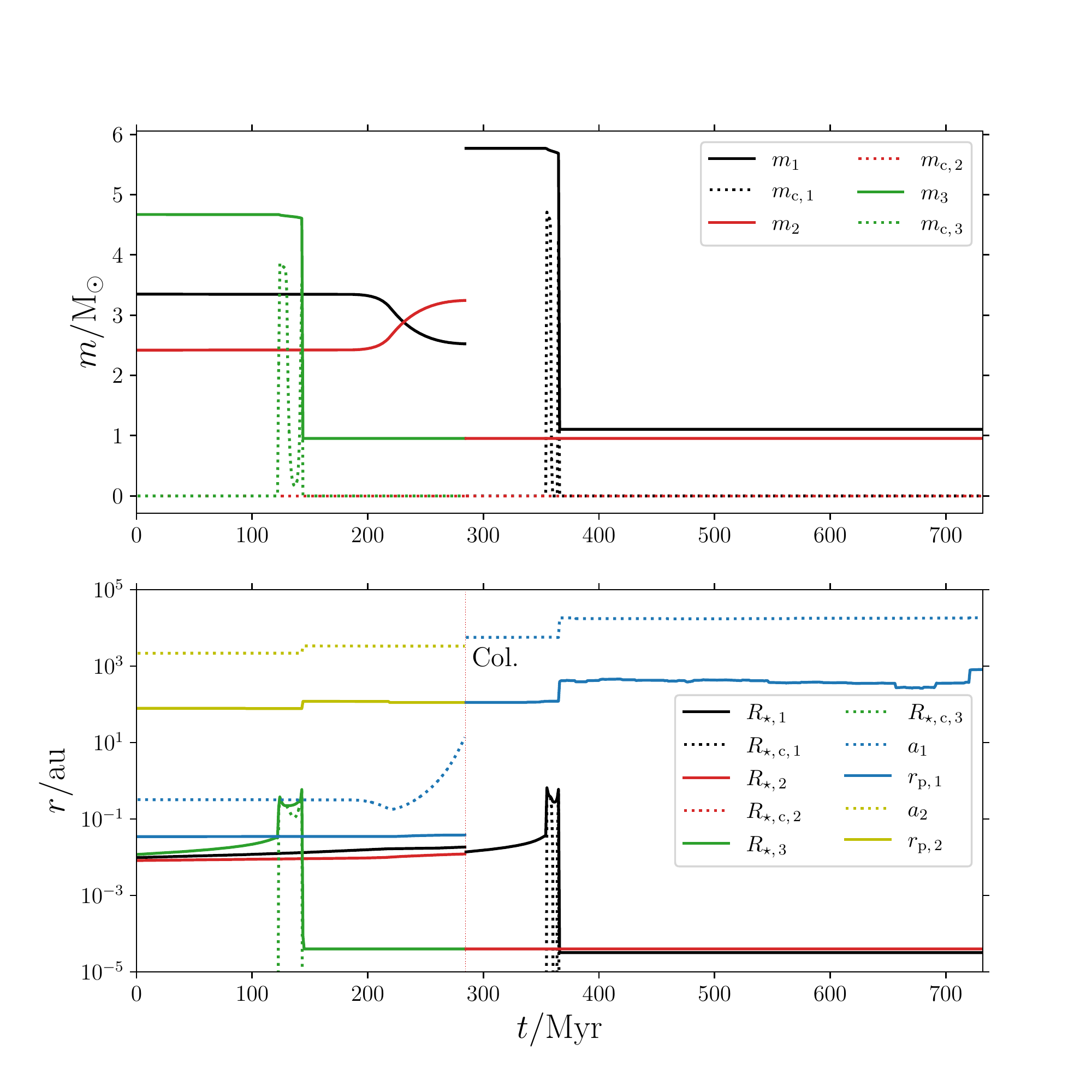}
}
\caption{Similar to \F~\ref{fig:ex1}, here for an example system (cf. \S\ref{sect:ex:2}) in which the TEDI is triggered by RLOF. Note that the inner orbit disappears after the collision at $\sim 285\,\myr$. }
\label{fig:ex2}
\end{figure}

\begin{table*}
\begin{center}
\begin{tabular}{ccccccccccccccc}
\toprule 
\S & $m_1$ & $m_2$ & $m_3$ & $a_1$ & $a_2$ & $e_1$ & $e_2$ & $i_1$ & $i_2$ & $\omega_1$ & $\omega_2$ & $\Omega_1$ & $\Omega_2$ & $i_\mathrm{rel}$ \\
\midrule
\ref{sect:ex:1} & 1.318 & 1.220 & 0.749 & 46.200 & 9631.601 & 0.699 & 0.845 & 1.942 & 1.480 & 0.068 & 0.100 & 1.872 & 4.013 & 2.133\\
\ref{sect:ex:2} & 3.348 & 2.420 & 4.670 & 0.320 & 2175.014 & 0.892 & 0.964 & 0.977 & 0.904 & 3.085 & 5.012 & 0.959 & 0.970 & 0.074 \\
\bottomrule
\end{tabular}
\end{center}
\caption{Initial conditions for the examples presented in \S\ref{sect:ex}. Masses $m_i$ are in units of $\msun$, and semimajor axes $a_i$ are in units of $\au$. The orbital angles (inclinations $i_i$, arguments of periapsis $\omega_i$, and longitudes of the ascending node $\Omega_i$) are expressed in radians. }
\label{table:exICs}
\end{table*}

\subsection{Traditional TEDI}
\label{sect:ex:1}
First, we consider a `traditional' TEDI scenario in which dynamical instability is triggered by mass loss from stellar winds. In Fig~\ref{fig:ex1_mob}, we show mobile diagrams \citep{1968QJRAS...9..388E} at key points in the evolution. The masses and orbital properties and stellar radii are shown as a function of time in \F~\ref{fig:ex1}. The most massive star (initially $m_1 \simeq 1.3 \, \msun$) evolves to a WD on a $\sim 5 \,\gyr$ time-scale. This widens both the inner and outer orbits, making the triple more compact (smaller $a_2/a_1$); however, note that fly-bys also play a role, in particular by increasing the outer orbital eccentricity (i.e., decreasing $r_\mathrm{p,\,2}$). Dynamical instability is triggered at $t\sim 5.5\,\gyr$, around which time the secondary star is also evolving and losing mass, further widening the orbits. Eventually, around $t\sim 5.7 \, \gyr$, the primary (now a WD) and the secondary (now a giant star) physically collide, triggering CE evolution and leaving a new giant star, which quickly evolves into a WD. Also as a result of dynamical instability, the tertiary star becomes unbound from the inner binary merger remnant.

\subsection{RLOF-induced TEDI}
\label{sect:ex:2}
In the second example (refer to Figs~\ref{fig:ex2_mob} and \ref{fig:ex2}), the TEDI is not triggered primarily by stellar wind mass loss, but by mass transfer. The inner binary is initially tight (semimajor axis $a_1 \sim 0.3 \, \au$) and eccentric ($e_1 \sim 0.89$). The tertiary star is the most massive and evolves first, consequently widening the outer orbit, whereas the inner orbit is unaffected. As the primary star expands on the MS, RLOF is triggered at $t \simeq 166 \, \myr$ (note that ZLK oscillations are unimportant, since $a_2(1-e_2) \gg a_1$, and $i_\mathrm{rel}$ is small). Initially, the inner orbit shrinks since $m_1>m_2$ (cf. the blue dotted line in the bottom panel of \F~\ref{fig:ex2}). After the inner binary mass ratio reaches unity at $t\simeq 220\,\myr$, the inner orbit expands, keeping the periapsis distance approximately constant (hence, $e_1$ increases). At $t\simeq 284\,\myr$, the expanding inner orbit triggers dynamical instability, and shortly afterwards, the inner binary (still consisting of two MS stars) merges\footnote{At $t\simeq 284.45\,\myr$, the now less massive primary star fills its Roche lobe again around the now more massive secondary; this can be understood by noting that the primary is still larger than the secondary (cf. \F~\ref{fig:ex2}), whereas its Roche lobe radius is smaller.}. A new binary system is formed with the merger remnant and the tertiary star, which is now a WD. As the merger remnant evolves to another WD, the orbit further widens and becomes more susceptible to fly-bys.

\section{Population synthesis: methods}
\label{sect:popmeth}
\subsection{Monte Carlo sampling}
\label{sect:popmeth:MC}
For our population synthesis calculations, we adopt a similar approach to that of \pk~to generate a population of initial triples using Monte Carlo-style sampling. We emphasize that many uncertainties exist in the initial distributions of triple and higher-order multiple systems, especially among systems with massive primaries. Here, instead of adopting a wide range of initial distributions, we choose to focus on one particular distribution and utilise our computational resources to investigate the impact of fly-bys. 

The `primary' star mass $m_1$ is sampled from a Kroupa distribution \citep{2001MNRAS.322..231K,2002Sci...295...82K}, $\mathrm{d}N/\mathrm{d}m_1 \propto m^{-2.35}$ for $1\,\msun<m_1<150\,\msun$. The secondary stellar mass $m_2$ is sampled from $q_1 \equiv m_2/m_1$, where $q_1$ is uniformly distributed between 0 and 1. The tertiary stellar mass $m_3$ is computed from $q_2 \equiv m_3/(m_1+m_2)$, where $q_2$ is also distributed uniformly between 0 and 1. We reject any sampled mass $m_2$ and $m_3$ if it is less than $0.08 \, \msun$, approximately the limit of central H burning. Note that our approach allows for the tertiary star to be more massive than the primary star (the star for which the mass was sampled directly from the IMF) in the inner binary, although not more massive than the total mass of the inner binary. Observations of triples indeed show that the tertiary star is the most massive star in some cases \citep[e.g.,][]{2010yCat..73890925T,2014MNRAS.438.1909D}.

Subsequently, the inner and outer orbital semimajor axes are sampled, where we adopt a lognormal distribution in the orbital period if the primary stellar mass is $m_1<3\,\msun$ \citep{1991A&A...248..485D,2010ApJS..190....1R}, and flat in log $a_i$ if $m_1 > 3 \, \msun$ \citep{2007ApJ...670..747K}. The sampled semimajor axes have the range $10^{-3} \, \au<a_i<10^5\,\au$, where the upper limit is motivated by the fact that much wider orbits are dissociated in the Galactic potential \citep[e.g.,][]{2018MNRAS.474.4412F}. The lower limit is to ensure the tightest possible systems, and in practice the actual lowest $a_i$ is $\gtrsim 10^{-2} \, \au$ (see below, and \F~\ref{fig:ICs}). 

Observations show that the distributions of orbital eccentricities in binary systems are on average less eccentric than a thermal distribution, $\mathrm{d} N/\mathrm{d}e_i \propto e_i$ \citep[e.g.,][]{2010ApJS..190....1R,2012Sci...337..444S,2013ARA&A..51..269D,2017ApJS..230...15M}. A thermal eccentricity distribution is expected for binaries in equilibrium embedded in stellar systems \citep{1919MNRAS..79..408J}, although tidal and pre-MS evolution likely affect the initial eccentricity distribution, especially for tighter binaries \citep[e.g.,][]{2018ApJ...854...44M,2019ApJ...872..165G}. The eccentricity distributions are less well constrained for higher-order multiplicity systems such as triples and quadruples, especially in the high mass range. Here, we adopt a Gaussian distribution for both inner and outer orbital eccentricities with a mean of $\mu=0.4$ and standard deviation of $\sigma=0.4$, as suggested  by \citet{2013ARA&A..51..269D} as a reasonable approximation to the observational data. We remark that our effective eccentricity distributions are modified by stability and non-Roche lobe filling requirements (see below).

The orbital angles (inclinations $i_i$, arguments of periapsis $\omega_i$, and longitudes of the ascending node $\Omega_i$, for the inner and outer orbits) are sampled assuming random orbital orientations, i.e., we adopt uniform distributions in $\cos i_i$, $\omega_i$, and $\Omega_i$. For simplicity, we ignore the recent indications that more tight triples are likely to be more coplanar, and wider triples more isotropically mutually oriented \citep{2017ApJ...844..103T}, and given that the mutual inclination distributions in triples are still uncertain, especially for massive triples. 

We reject a sampled system if it is initially unstable according to the criterion of \citeauthor{2001MNRAS.321..398M} (\citeyear{2001MNRAS.321..398M}; cf. \eq~\ref{eq:dynstab}). We also reject systems in which either inner binary stars would immediately fill their Roche lobe around each other at periapsis, where we use the fit of \citet{1983ApJ...268..368E} to calculate the Roche lobe radius evaluated at the periapsis distance $a_1(1-e_1)$, and estimate the initial MS radii\footnote{This estimate only applies to the sampling of initial systems; the radii are calculated more accurately in the simulations based on the \textsc{SSE} tracks \citep{2000MNRAS.315..543H}.} as $R_\star = R_\odot \, (m_i/\msun)^{0.7}$. 

The resulting cumulative distributions of the semimajor axes and eccentricities are shown in \F~\ref{fig:ICs}. As is customary in triple population synthesis studies \citep[e.g.][]{2007ApJ...669.1298F,2013MNRAS.430.2262H}, the dynamical stability criterion ensures that the inner semimajor axis distribution is peaked at a much smaller value compared to the outer semimajor axis, despite that both $a_1$ and $a_2$ are sampled from the same underlying distributions. Furthermore, the stability criterion disfavours high outer orbital eccentricities. High initial inner orbital eccentricities are also disfavoured, although this can be attributed to the requirement of the inner binary stars not filling their Roche lobes initially. 

We simulate each successfully sampled system for a duration of $t_\mathrm{end}=10\,\gyr$. We include two sets, each consisting of $N_\mathrm{MC}=10^5$ systems: one in which fly-bys are included, and one in which fly-bys are excluded. Here, fly-bys are taken into account using the default parameters in \mse, i.e., we adopt a local stellar density of $n_\star=0.1\,\pc^{-3}$, a relative velocity dispersion of $30 \, \kms$, and an encounter sphere radius of $R_\mathrm{enc}=10^5\,\au$ \citep[e.g.,][]{2008gady.book.....B}. The perturber masses follow the \citet{1993MNRAS.262..545K} distribution (taking into account gravitational focussing). The computational wall time per system was limited to 10 hr; the latter was exceeded for $\approx 4\%$ of sampled systems, but we do not expect this to significantly affect our results. 

\begin{figure}
\iftoggle{ApJFigs}{
\includegraphics[width=\linewidth]{ICs_orbits_CDF_run11}
}{
\includegraphics[width=\linewidth]{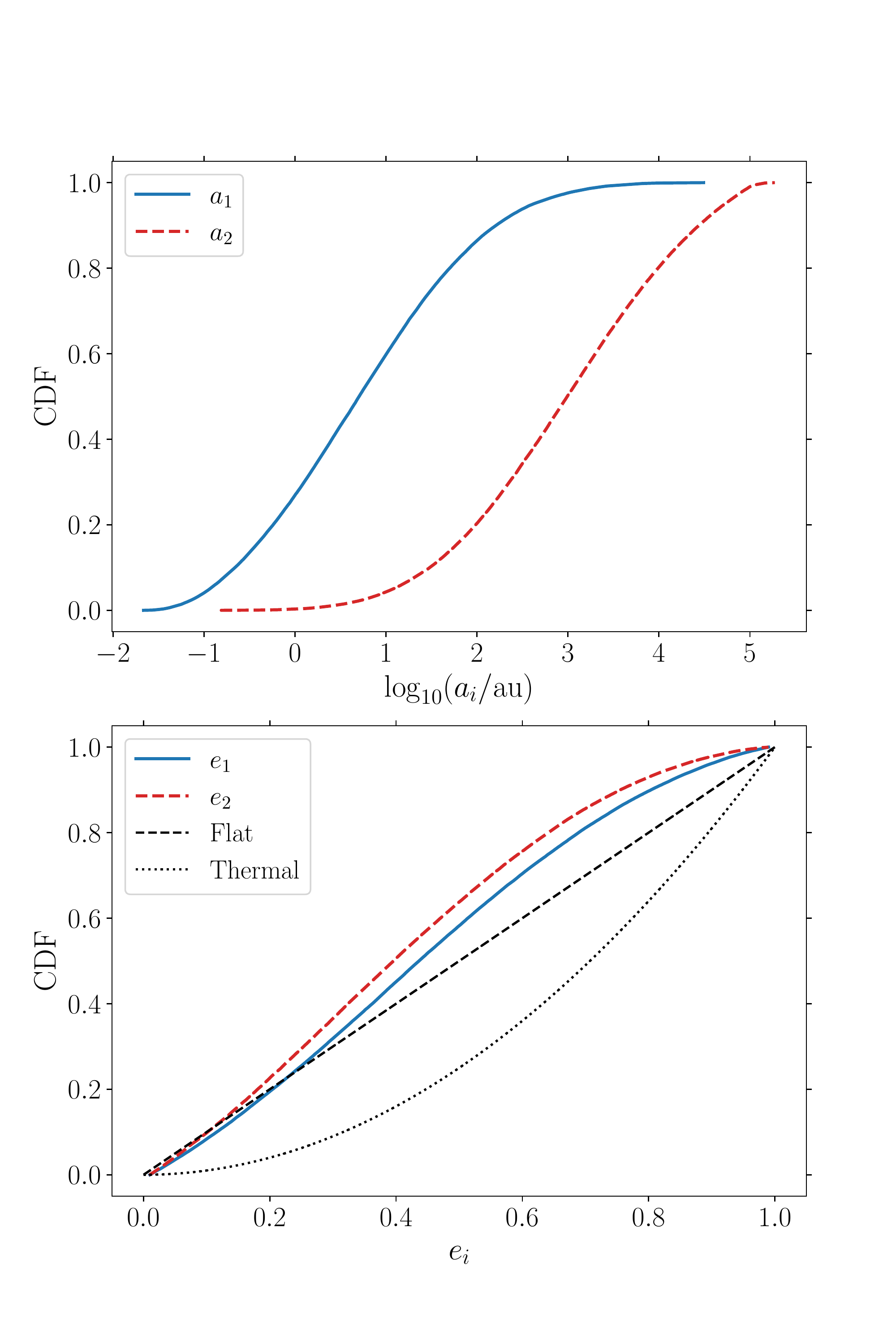}
}
\caption{Cumulative distributions of the initial semimajor axes (top panel) and eccentricities (bottom panel) for the population synthesis simulations (cf. \S\ref{sect:popmeth}). Blue solid (red dashed) lines correspond to the inner (outer) orbit. In the bottom panel, the black dashed and dotted lines show flat and thermal distributions, respectively.}
\label{fig:ICs}
\end{figure}

\subsection{Classification of TEDI outcomes}
\label{sect:popmeth:ch}
In the population synthesis calculations, we define TEDI systems as those in which a dynamical instability occurred (\eq~\ref{eq:dynstab}) at any time in the evolution, and/or the semisecular regime was entered (\eq~\ref{eq:semisec}). We remark that semisecular systems are not unstable in the strictest sense, but often lie on the boundary between being stable and unstable. We here choose to include semisecular systems in our definition of TEDI systems (this is also in line with the fact that \mse~switches to direct $N$-body integration in either cases of equations~\ref{eq:dynstab} and \ref{eq:semisec}).

We classify the following outcomes for a system in which TEDI was triggered.
\begin{enumerate}
\item Head-on collision between the two stars in the inner binary system. This includes `clean' collisions involving more compact stars, as well as collisions involving giant stars with extended envelopes (in the latter case, the code could invoke CE evolution).
\item A head-on collision involving the tertiary star and one star in the inner binary system.
\item Stable: after the instability was triggered and has passed (the system is now dynamically stable again), the triple remains with the same structure, although the orbits could have changed substantially.
\item The tertiary star escapes, leaving the inner binary system which continues as an isolated bound binary.
\item Following an exchange interaction, the tertiary star becomes a bound pair with one of the stars in the inner binary, whereas the other inner binary component escapes.
\item All three stars become unbound from each other and escape.
\end{enumerate}

\section{Population synthesis: results}
\label{sect:results}

\begin{table}
\begin{center}
\begin{tabular}{lcc}
\toprule 
Outcome & \multicolumn{2}{c}{Fraction} \\
& Fly-bys & No Fly-bys \\
\midrule
All mergers & $0.534 \pm 0.002$ & $0.533 \pm 0.002$ \\
CE & $0.484 \pm 0.002$ & $0.484 \pm 0.002$ \\
Clean collision & $0.144 \pm 0.001$ & $0.138 \pm 0.001$ \\
\midrule
Dynamical inst. (DI) & $0.0397 \pm 0.0006$ & $0.0302 \pm 0.0005$ \\
Secular breakdown (SB) & $0.0200 \pm 0.0004$ & $0.0163 \pm 0.0004$ \\
TEDI (DI or SC) & $0.0526 \pm 0.0007$ & $0.0419 \pm 0.0006$ \\
\bottomrule
\end{tabular}
\end{center}
\caption{Outcome fractions of various main channels in the population synthesis calculations (\S\ref{sect:results}). We distinguish between all collisions (which can include CE evolution and/or pure collisions), dynamical instability according to \eq~(\ref{eq:dynstab}), and secular breakdown according to \eq~(\ref{eq:semisec}). We consider a system to be a TEDI system if either dynamical instability (DI) or secular breakdown (SB) occurred during the evolution. Results are shown for simulations with and without fly-bys taken into account. Statistical error bars are given based on Poisson statistics (i.e., for an outcome $X$, the fractional error is $\sqrt{N_X}/N_\mathrm{MC}$). Note that the error bars do not reflect systematic uncertainties. }
\label{table:gen_fractions}
\end{table}

\subsection{Overall fractions}
\label{sect:results:frac}
We begin by quoting in Table~\ref{table:gen_fractions} the fractions of several major outcomes in our population synthesis simulations, not limiting to those systems undergoing TEDI (with or without fly-bys included). In $\sim 54\%$ of our simulations, stars collide during any point in the simulation. Here, our definition of `collision' includes both `clean' collisions (with the instantaneous separation being less than the sum of the radii and involving stars without extended envelopes such as MS stars and compact objects), and CE evolution (which can be triggered after a star fills its Roche lobe around a companion, or when at least one of two physically-colliding stars has an extended envelope). Many of these `collisions' involve CE evolution, whereas clean collisions are less common (still, their fractions of $\sim 14\%$ are significant). We remark that `all collisions' in Table~\ref{table:gen_fractions} also includes systems in which both CE and clean collisions occur, hence the `CE' and 'pure collision' fractions do not add up to the `all collisions' fractions. 

Dynamical instability (according to \eq~\ref{eq:dynstab}) occurs in $\sim 4\%$ of systems (with fly-bys included), whereas a secular breakdown (cf. \eq~\ref{eq:semisec}) occurs in $\sim 2\%$ of systems. We remind the reader that we consider either of these channels for identifying TEDI systems (cf. \S\ref{sect:popmeth:ch}). The overall TEDI fraction (with fly-bys) is $\sim5\%$ for all TEDI-related channels. In \S\ref{sect:results:rate}, we focus further on the different outcomes of TEDI systems in our simulations. 

Fly-bys do not significantly affect the total number of collisions in our simulations. However, the fraction of dynamically unstable systems is significantly higher, and increases from $\sim 3\%$ to $\sim 4\%$ when fly-bys are taken into account. The overall TEDI fraction increases from $\sim 4$ to $\sim5\%$, which is statistically significant. We remark, however, that the error bars quoted in Table~\ref{table:gen_fractions} are statistical errors, and hence do not take into account systematic errors due to, e.g., uncertainties in CE evolution, or the omission of tidal effects in the $N$-body simulations (cf. \S\ref{sect:meth:dyn}).

\subsection{Galactic event rates}
\label{sect:results:rate}
In this section, we estimate the Galactic rates of different outcomes of the TEDI in our simulations. The normalisation procedure is described first. 

\subsubsection{Normalisation}
\label{sect:results:rate:norm}
We assume a Galactic star formation rate of $R_\mathrm{SFR} = 1 \,\msun\,\yr^{-1}$ \citep{2010ApJ...710L..11R}. This formed mass per unit time is assumed to comprise of single, binary, and triple systems (here, we do not consider high-order systems such as quadruple systems, although dynamical instabilities can also be triggered in the latter, e.g., \citealt{2018MNRAS.478..620H}). The underlying initial mass function (IMF) is assumed to be that of \citet{2001MNRAS.322..231K,2002Sci...295...82K}, i.e., a broken power law with
\begin{align}
\label{eq:imf}
\frac{\mathrm{d}N}{\mathrm{d} m} \propto \left \{ 
\begin{array}{cc} 
m^{-1.3}, & 0.08\,\msun<m<0.5 \, \msun; \\
m^{-2.3}, & 0.5\,\msun<m<1\,\msun; \\
m^{-2.35}, & 1\,\msun<m<150\,\msun.
\end{array}
\right.
\end{align}
In our simulations, we limited the primary star mass to $m_1>1\,\msun$; \eq~(\ref{eq:imf}) then implies that the fraction of calculated systems compared to all systems is $f_\mathrm{calc} \simeq 0.09965$. 

Furthermore, the IMF \eq~(\ref{eq:imf}) implies that a population of $N_\mathrm{s}$ single stars has an average mass of $M_\mathrm{s} \equiv M_\mathrm{Kr} N_\mathrm{s}$, where $M_\mathrm{Kr} \simeq 0.5774 \, \msun$. Assuming flat mass ratio distributions, a population of $N_\mathrm{bin}$ binary stars has a total mass of $M_\mathrm{bin} \approx (1+\frac{1}{2} ) M_\mathrm{Kr} N_\mathrm{bin} = \frac{3}{2} M_\mathrm{Kr} N_\mathrm{bin}$. Also, assuming a flat distribution of the outer mass ratio $q_2 \equiv m_3/(m_1+m_2)$, a population of $N_\mathrm{tr}$ triple stars has a total mass of $M_\mathrm{tr} \approx (\frac{3}{2} + \frac{1}{2} \frac{3}{2} ) M_\mathrm{Kr} N_\mathrm{tr} = \frac{9}{4} M_\mathrm{Kr} N_\mathrm{tr}$. 

With the above assumptions, a population of single, binary, and triple stars has a total mass of
\begin{align}
\nonumber M_\mathrm{tot} & = M_\mathrm{s} + M_\mathrm{bin} + M_\mathrm{tr} \\
\nonumber &\approx \left (N_\mathrm{s} + \frac{3}{2} N_\mathrm{bin} + \frac{9}{4} N_\mathrm{tr} \right ) M_\mathrm{Kr} \\
 &= \left (1 + \frac{1}{2} \alpha_\mathrm{bin} + \frac{5}{4} \alpha_\mathrm{tr} \right) N_\mathrm{sys} M_\mathrm{Kr},
\end{align}
where $\alpha_\mathrm{s}$, $\alpha_\mathrm{bin}$, and $\alpha_\mathrm{tr}$ are the single, binary, and triple fractions, respectively ($\alpha_\mathrm{s} + \alpha_\mathrm{bin} + \alpha_\mathrm{tr} = 1$), and $N_\mathrm{sys}$ is the total number of stellar systems. We assume mass-independent fractions $\alpha_\mathrm{bin} = 0.6$, and $\alpha_\mathrm{tr} = 0.1$. 

In terms of $N_\mathrm{sys}$, the number of systems calculated in our simulations is given by $N_\mathrm{calc} = f_\mathrm{calc} \alpha_\mathrm{tr} N_\mathrm{sys}$. For a particular outcome $X$ with $N_X$ systems, the corresponding Galactic rate $R_X$ is then given by
\begin{align}
\nonumber R_X &= N_X \frac{R_\mathrm{SFR}}{M_\mathrm{tot}} = \frac{N_X f_\mathrm{calc} \alpha_\mathrm{tr} R_\mathrm{SFR}}{N_\mathrm{calc} \left (1 + \frac{1}{2} \alpha_\mathrm{bin} + \frac{5}{4} \alpha_\mathrm{tr} \right) M_\mathrm{Kr}} \\
&\simeq 0.01211 \, \yr^{-1} \, \left ( \frac{N_X}{N_\mathrm{calc}} \right )\left ( \frac{R_\mathrm{SFR}}{1 \, \msun \, \yr^{-1}} \right ),
\end{align}
where we substituted $R_\mathrm{SFR} = 1\,\msun\,\yr^{-1}$ for the numerical value; note that $N_X/N_\mathrm{calc} = f_X$ is simply the fraction of systems in the population synthesis calculations corresponding to outcome $X$.

\begin{table}
\begin{center}
\begin{tabular}{lcc}
\toprule 
Channel & \multicolumn{2}{c}{Galactic Rate ($10^{-4} \, \mathrm{yr^{-1}}$)} \\
& Fly-bys & No Fly-bys \\
\midrule
TEDI channels \\
\midrule
Collision (all) & $2.47 \pm 0.05$ & $2.12 \pm 0.05$ \\
~Collision (MS-MS) & $1.61 \pm 0.04$ & $1.43 \pm 0.04$ \\
~Collision (Giant-MS) & $0.59 \pm 0.03$ & $0.55 \pm 0.03$ \\
~Collision (WD-MS) & $0.20 \pm 0.02$ & $0.08 \pm 0.01$ \\
~Collision (WD-giant) & $0.03 \pm 0.01$ & $0.04 \pm 0.01$ \\
~Collision (WD-WD) & $0.04 \pm 0.01$ & $0.01 \pm 0.00$ \\
~Any collision w. tertiary & $0.06 \pm 0.01$ & $0.06 \pm 0.01$ \\
Tertiary unbound & $0.56 \pm 0.03$ & $0.40 \pm 0.02$ \\
Exchange \& single unbound & $1.01 \pm 0.03$ & $0.94 \pm 0.03$ \\
All unbound & $1.25 \pm 0.04$ & $0.44 \pm 0.02$ \\
\midrule
All collision channels & $64.6 \pm 0.3$ & $64.5 \pm 0.3$ \\
\bottomrule
\end{tabular}
\end{center}
\caption{Galactic rates for various outcomes of the TEDI according to the population synthesis calculations (\S\ref{sect:results}): collisions (making a distinction between all collisions, collisions involving specific types of stars, and any collisions involving the tertiary star), the tertiary star becoming unbound, leaving the inner binary, an exchange interaction in which the tertiary star and an inner binary component remain, and the other inner binary component escapes, and, lastly, all three stars becoming unbound. Results are shown for the simulations with and without fly-bys. The bottom row gives the total overall collision rate in the simulations, which includes but is not limited to TEDI-induced collisions. Error bars indicate statistical Poisson errors.}
\label{table:rates}
\end{table}

\subsubsection{Results}
\label{sect:results:rate:res}
The Galactic rates of various outcomes of the TEDI in our simulations with the normalisation as described in \S\ref{sect:results:rate} are shown in Table~\ref{table:rates}. We distinguish between TEDI-induced collisions (making a distinction between all collisions, collisions per stellar type, and any collisions involving the tertiary star), the tertiary star becoming unbound, an exchange interaction in which one of the inner binary escapes, and all three stars becoming unbound. Results are shown for simulations in which fly-bys were included and excluded. 

We remind the reader that our focus is on systems that become dynamically unstable, and TEDI systems were defined accordingly (cf. \S\ref{sect:popmeth:ch}). For reference, the bottom row shows the total Galactic collision rate in our simulations from {\it all triple channels}, irrespective of origin. The total collision rate includes, but is not limited to TEDI-induced collisions; e.g., mergers following `standard' binary evolution (possibly enhanced by secular evolution) are also included. The TEDI-induced collision rates are significantly lower than the overall rates, which can be understood from the fact that the overall fraction of dynamically unstable or semisecular systems is only $\sim 5\%$ (cf. Table~\ref{table:gen_fractions}). Moreover, the overall collision rate is dominated by triples in which binary interactions such as mass transfer and CE evolution in the inner orbit play a major role, and which do not require an instability phase to occur (although they can be enhanced by stable secular evolution). 

Our total Galactic collision rate (for all channels) is roughly on the order of $10^{-2}\,\yr^{-2}$, which can be compared to the estimated observed Galactic rate \citep{2014MNRAS.443.1319K} of $\sim 0.5 \, \yr^{-1}$ for all stellar mergers (both clean collisions and CE evolution). This shows that triples contribute a small but non-negligible fraction of all stellar collisions in the Galaxy, though a more detailed comparison is beyond the scope of this paper.

Among TEDI systems, collisions dominate. Most collisions involve two MS stars, and typically originate from systems that are initially only marginally stable; dynamical instability occurs quickly, and collisions occur when the stars are still on the MS (cf. \S\ref{sect:results:st} and \S\ref{sect:results:t}). The total rate of TEDI-related collisions is $R_\mathrm{col} \sim 2 \, \times 10^{-4}\, \yr^{-1}$. This rate is quite consistent with the value $R_\mathrm{col,\,PK12} = 1.2 \times 10^{-4}\, \yr^{-1}$ estimated by \pk, in particular when MS-MS collisions (most of which were not taken into account by \pk) are excluded.

Among TEDI-collisions involving more evolved stars (beyond the MS), giant-MS collisions are most frequent, followed by WD-MS collisions. In rare cases (Galactic event rates $\sim 10^{-6}\,\yr^{-1}$), two WDs collide, potentially producing SNe Ia \citep[e.g.,][]{2009ApJ...705L.128R,2010MNRAS.406.2749L,2011A&A...528A.117P,2016ApJ...822...19P}\footnote{We note that in \citet{2011A&A...528A.117P}, the definition of `collision' included dynamically unstable RLOF.}. 

Nearly all TEDI-induced collisions occur between the two inner binary stars. Only $\sim 2\%$ of TEDI-collisions involve the tertiary star (fly-bys included). This is in contrast to dynamical scattering involving triple-single, triple-binary, and binary-binary interactions, which favor more equal collision probabilities \citep{2016MNRAS.456.4219A}. 

Unbound stars and exchange interaction channels are each less common than all collisions. When MS-MS collisions are excluded, however, the collision rates are comparable to that of the unbound and exchange interaction channels. 

Fly-bys systematically increase the rates of all TEDI channels, with an increase of $\sim 17\%$ for all collisions. The effect on the total collision rate is not significant, however.

\begin{table}
\begin{center}
\begin{tabular}{cl}
\toprule 
$k_i$ & Description \\
\midrule
0 & Main sequence ($m_i\lesssim0.7 \, \msun$) \\
1 & Main sequence ($m_i\gtrsim0.7 \, \msun$) \\
2 & Hertzsprung gap (HG) \\
3 & Red giant branch (RGB) \\
4 & Core helium burning (CHeB) \\
5 & Early asymptotic giant branch (EAGB) \\
6 & Thermally pulsing AGB (TPAGB) \\
7 & Naked helium star MS (He MS) \\
8 & Naked helium star Hertzsprung gap (He HG) \\
9 & Naked helium star giant branch (He GB) \\
10 & Helium white dwarf (He WD) \\
11 & Carbon-oxygen white dwarf (CO WD) \\
12 & Oxygen-neon white dwarf (ONe WD) \\
13 & Neutron star (NS) \\
14 & Black hole (BH) \\
15 & Massless remnant \\
\bottomrule
\end{tabular}
\end{center}
\caption{\small Description of the different stellar types used in \mse, reproduced from \citet{2000MNRAS.315..543H}.}
\label{table:st}
\end{table}

\begin{figure}
\iftoggle{ApJFigs}{
\includegraphics[width=1.1\linewidth,trim = 10mm 0mm 0mm 0mm]{TEDI_stellar_types_col}
\includegraphics[width=1.1\linewidth,trim = 10mm 0mm 0mm 0mm]{all_stellar_types_CE_col}
}{
\includegraphics[width=1.0\linewidth,trim = 10mm 0mm 0mm 0mm]{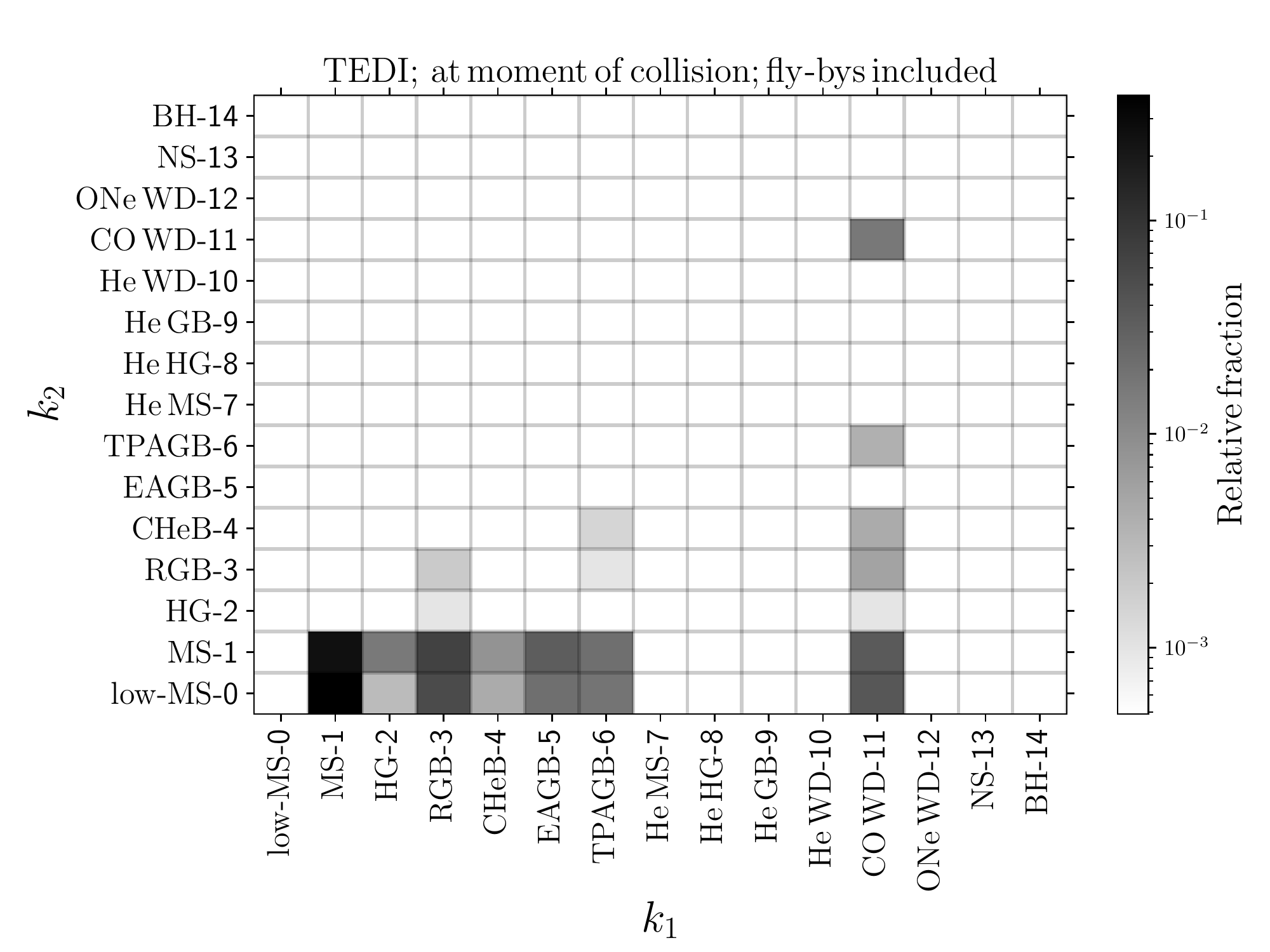}
\includegraphics[width=1.0\linewidth,trim = 10mm 0mm 0mm 0mm]{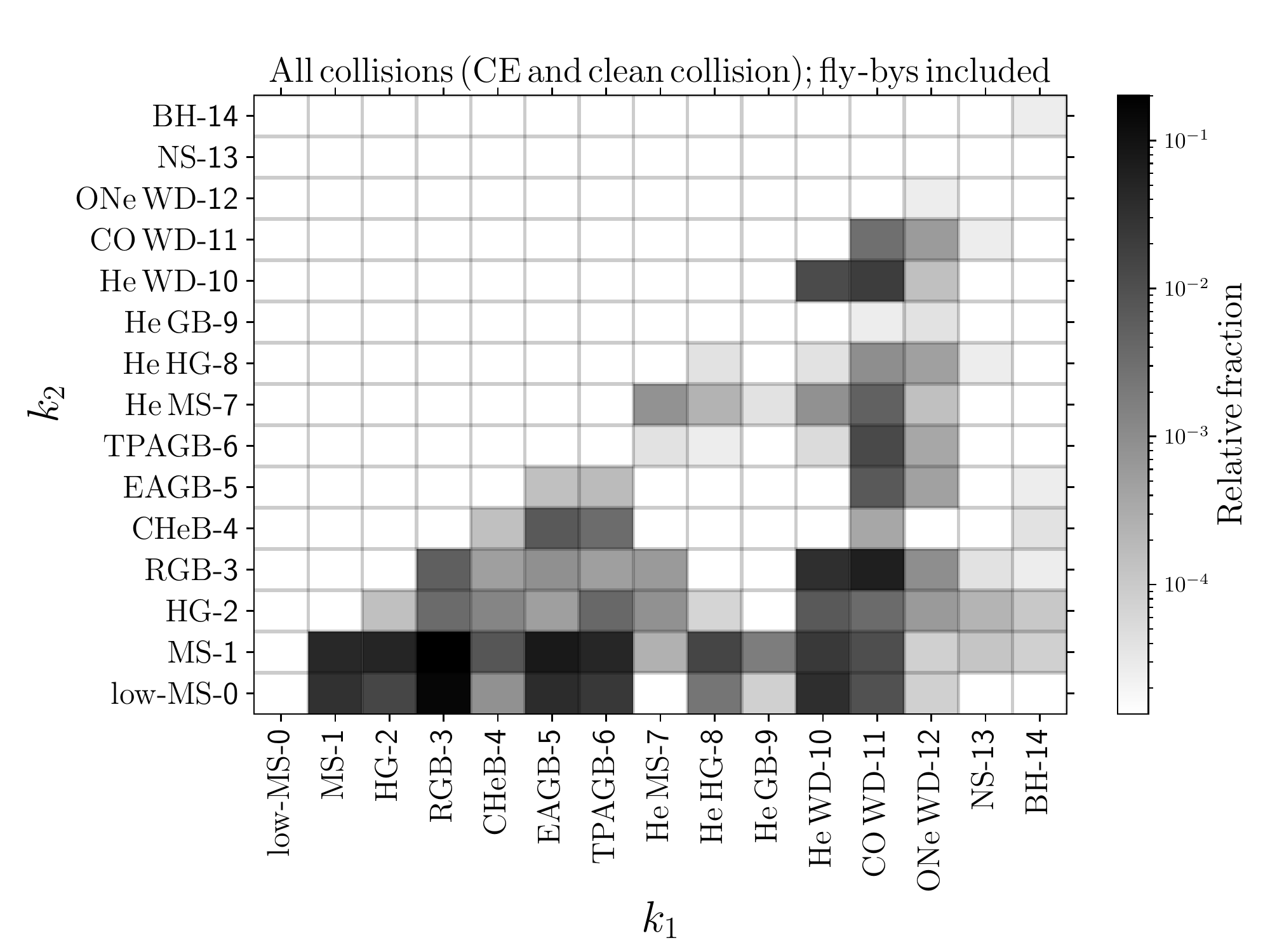}
}
\caption{Top panel: stellar types of colliding objects following a TEDI in our simulations with fly-bys included (refer to Table~\ref{table:st} for the definitions of the stellar types). The greyshade encodes the relative fraction of different combinations of stellar types, compared to all TEDI-induced collisions. Bottom panel: stellar types of colliding objects in {\it any} channel in our simulations (including, but not limited to TEDI-induced channels; fly-bys included). Note that the majority of collisions do not involve the TEDI (cf. Table~\ref{table:gen_fractions}). }
\label{fig:st_col}
\end{figure}

\begin{figure}
\iftoggle{ApJFigs}{
\includegraphics[width=1.1\linewidth,trim = 10mm 0mm 0mm 0mm]{TEDI_stellar_types_ad}
\includegraphics[width=1.1\linewidth,trim = 10mm 0mm 0mm 0mm]{TEDI_stellar_types_RLOF}
}{
\includegraphics[width=1.0\linewidth,trim = 10mm 0mm 0mm 0mm]{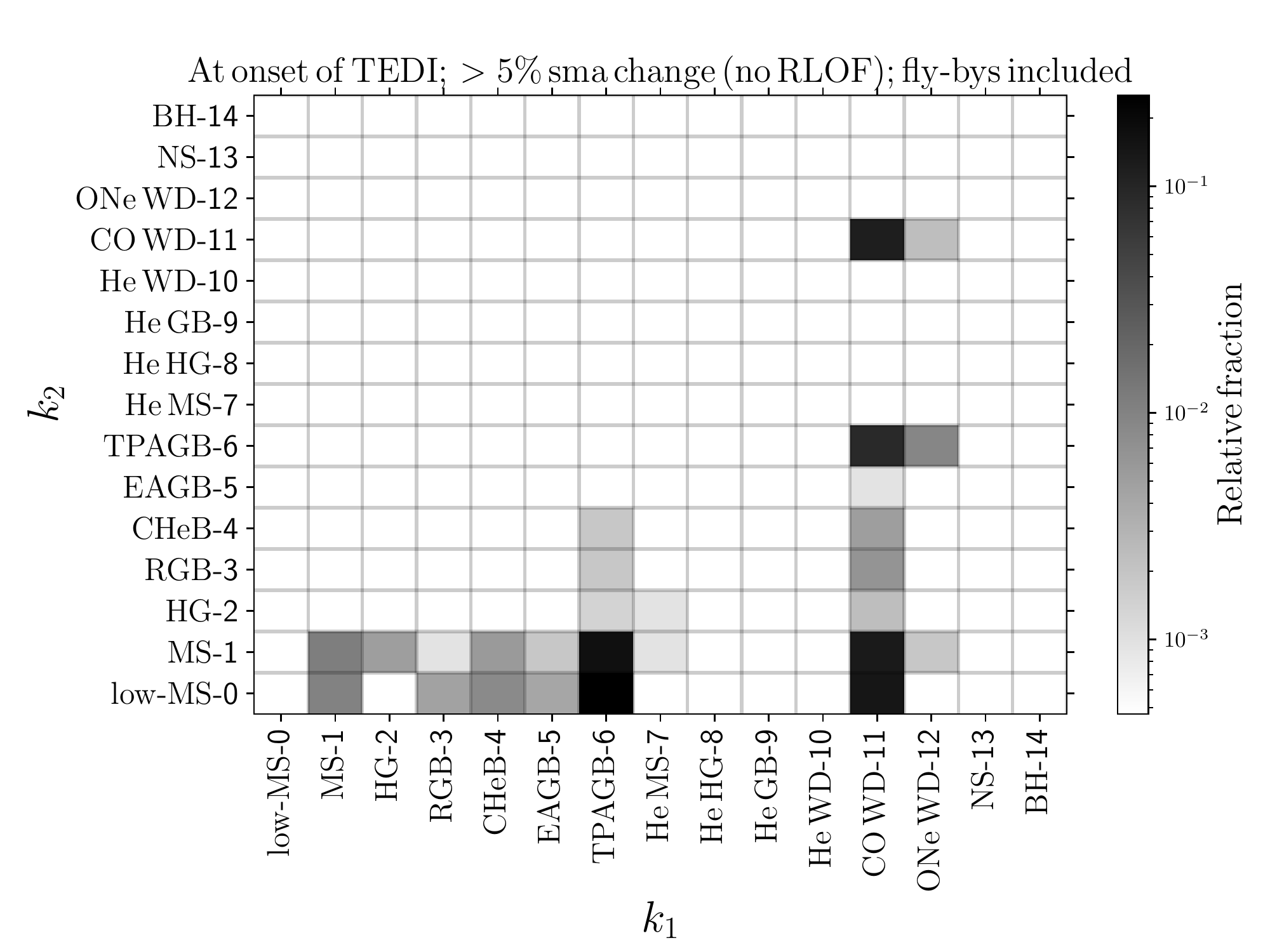}
\includegraphics[width=1.0\linewidth,trim = 10mm 0mm 0mm 0mm]{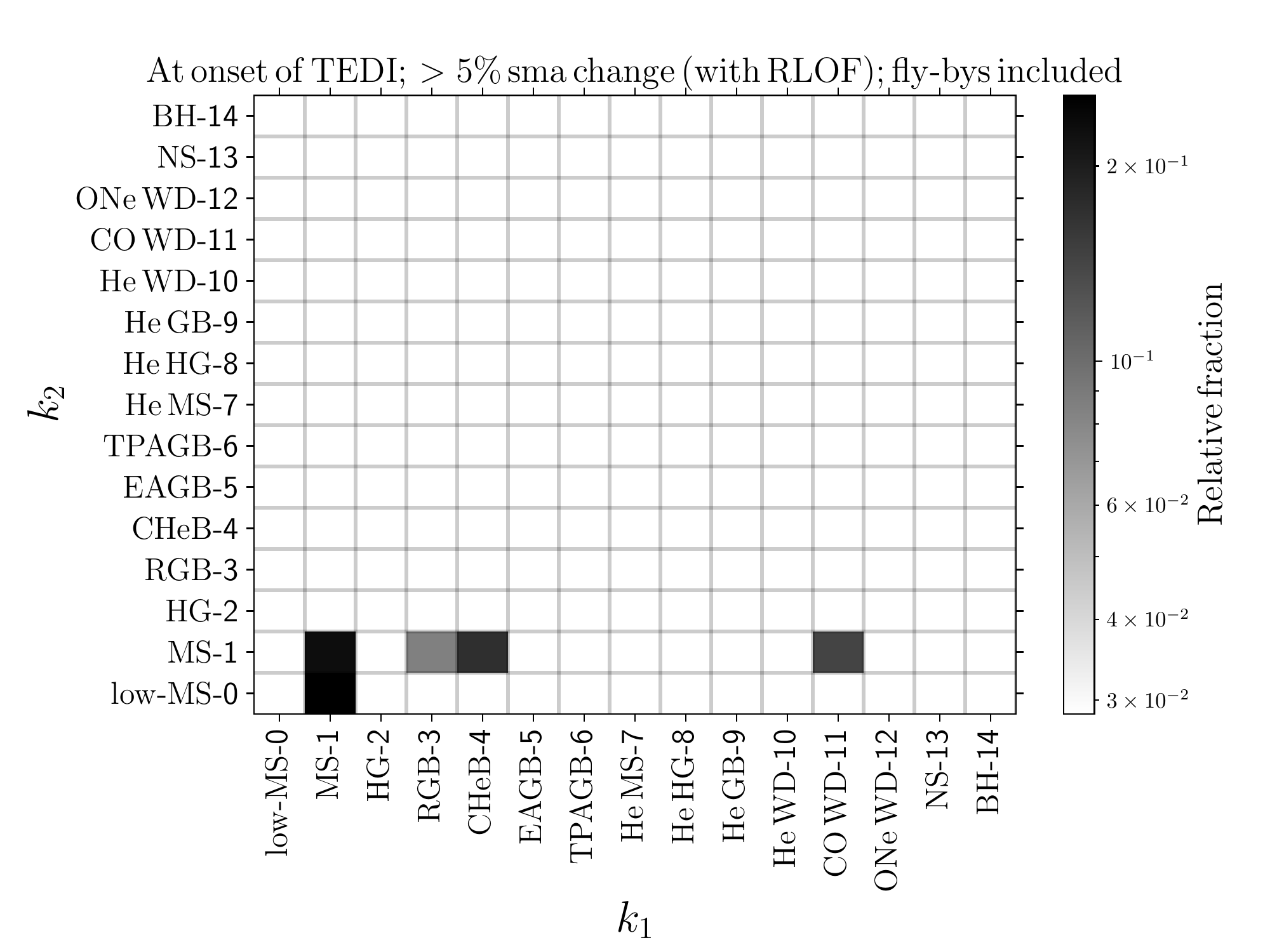}
}
\caption{Top panel: relative fractions divided into stellar types of the two stars in the inner binary when TEDI was triggered which involved an increase in the inner binary semimajor axis by at least 5\%, and where no RLOF was involved. The inner binary expansion in this case is attributed to adiabatic wind mass loss (the `traditional' TEDI channel). Bottom panel:  similar as the top panel, but now showing the relative fractions when TEDI was triggered by an increase in the inner binary semimajor axis by at least 5\% and which did involve RLOF. Fly-bys were included for both panels.}
\label{fig:st_ad_RLOF}
\end{figure}

\subsection{Stellar types}
\label{sect:results:st}
\subsubsection{At moment of collision}
\label{sect:results:st:col}
In \S\ref{sect:results:rate:res}, TEDI-induced collision rates were shown for a limited number of stellar type combinations. Here, we disentangle the different collision occurrences in more detail. The top panel in \F~\ref{fig:st_col} shows a stellar type `chessboard plot', i.e., the relative fractions (encoded in greyscale) in the $(k_1,k_2$) plane of collisions involving stars of types $k_1$ and $k_2$  following a TEDI in our simulations (with fly-bys included; refer to Table~\ref{table:st} for the definition of the stellar types). The greyshade encodes the relative fraction of different combinations of stellar types, compared to all TEDI-induced collisions. 

For reference, we also show in the bottom panel of \F~\ref{fig:st_col} the stellar type fractions for {\it all} collision channels in our simulations. We remark that TEDI-induced collisions only constitute a small fraction of all collisions; the bottom panel of \F~\ref{fig:st_col} indeed shows a significantly larger variety of stellar type combinations. 

To further understand the origin of TEDI collisions, we show in the top panel of \F~\ref{fig:st_ad_RLOF} relative fractions in the $(k_1,k_2$) plane for systems when TEDI was triggered which involved an increase in the inner binary semimajor axis (at the time of instability, compared to the initial semimajor axis) by at least 5\%, and for which no RLOF was involved. In these systems, the inner orbital expansion can be attributed to adiabatic wind mass loss (i.e., the `traditional' TEDI channel). We also show in the bottom panel of \F~\ref{fig:st_ad_RLOF} similar relative fractions for TEDI systems in which the inner orbit expanded by at least 5\%, but which also involved RLOF between the time of formation and the onset of instability. 

The top panel in \F~\ref{fig:st_ad_RLOF} shows that the `traditional' TEDI systems are likely to involve more evolved stars ($k_i>1$), which is expected since this channel is associated with adiabatic mass loss from evolving stars. The most common adiabatic TEDI collision systems involve giant stars, in particular thermally pulsing AGB stars ($k_1 = 6$) that are rapidly losing mass as they are evolving to become WDs. This is consistent with \pk, who found that most collisions in their simulations are associated with AGB stars. 

In contrast, TEDI systems driven by RLOF (cf. the bottom panel of \F~\ref{fig:st_ad_RLOF}) show a smaller range in stellar types at the onset of instability. A large fraction of these systems occur already during the MS, i.e., a tight inner binary fills its Roche lobe during the MS; as soon as the donor becomes more massive than the accretor, the inner binary expands, eventually driving triple dynamical instability (see \S\ref{sect:ex:2}). This channel is also possible with more evolved stars, most notably core He burning-MS and WD-MS systems. 

\begin{figure}
\iftoggle{ApJFigs}{
\includegraphics[width=1.1\linewidth,trim = 20mm 10mm 0mm 0mm]{TEDI_stellar_types_col_single_outcomes}
}{
\includegraphics[width=1.1\linewidth,trim = 20mm 10mm 0mm 0mm]{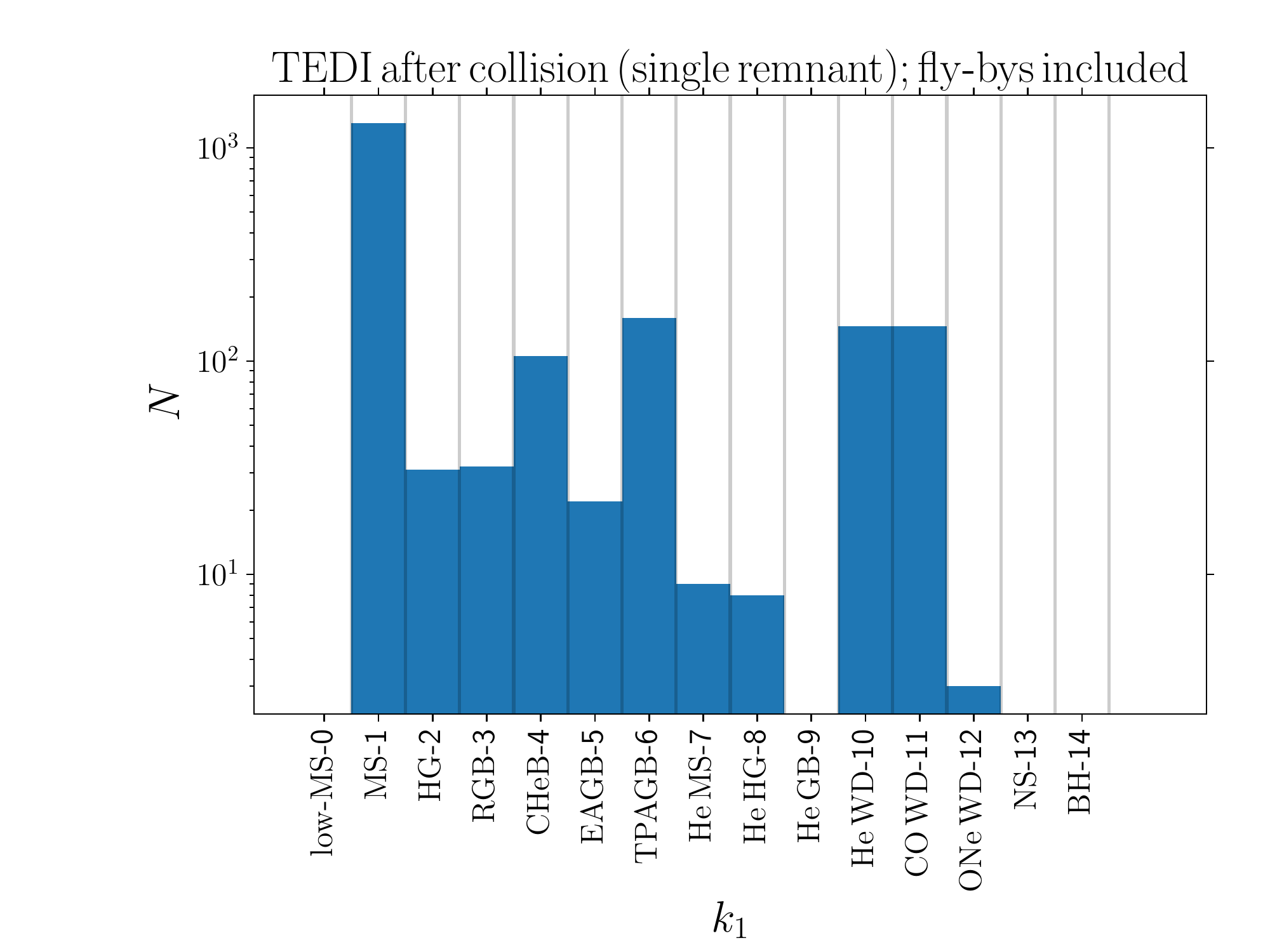}
}
\caption{Histogram of the stellar types of TEDI-induced merger remnants (with only one remnant resulting from the two colliding stars), in the simulations with fly-bys included. }
\label{fig:st_col_single_out}
\end{figure}

\begin{figure}
\iftoggle{ApJFigs}{
\includegraphics[width=1.0\linewidth,trim = 10mm 10mm 0mm 0mm]{TEDI_stellar_types_col_double_outcomes}
}{
\includegraphics[width=1.0\linewidth,trim = 10mm 10mm 0mm 0mm]{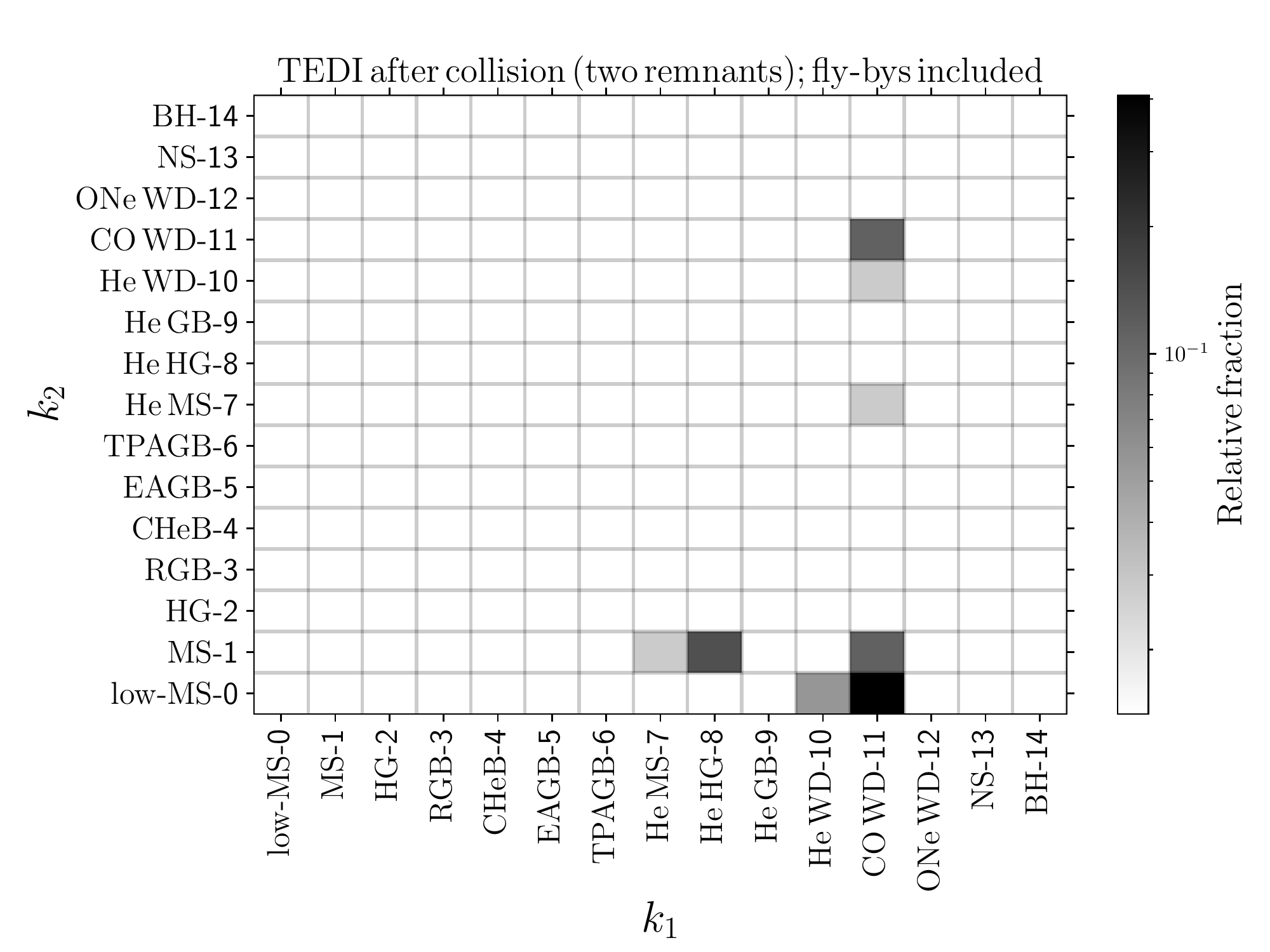}
}
\caption{Relative fractions of TEDI-induced collisions resulting in two merger remnants from the two colliding stars, divided into the stellar types of the merger remnants (fly-bys included).}
\label{fig:st_col_double_out}
\end{figure}

\subsubsection{Collision outcomes}
\label{sect:results:st:out}
After a collision following the TEDI, \mse~continues the evolution; the merger remnant is usually a single star, or a tight binary (rare exceptions being colliding sufficiently massive WDs, would could lead to a SNe Ia explosion, destroying both stars). Note that \mse~follows (highly simplified) prescriptions for CE evolution in the event that collisions involve a giant star with an extended envelope, and a more compact companion (see \citealt{2002MNRAS.329..897H,2021MNRAS.502.4479H}). In  \Fs~\ref{fig:st_col_single_out} and \ref{fig:st_col_double_out}, we show the corresponding histograms/relative fractions of the stellar types for the single and double remnant outcome cases, respectively. 

In the case of a single remnant (\F~\ref{fig:st_col_single_out}), the most likely outcome is a new MS, since collisions frequently involve two MS stars, which are assumed to merge into another MS star. Other likely single merger remnant outcomes are (sub)giant stars ($k_1\in\{2,3,5,6\}$), core He burning stars ($k_1=4$), and WDs ($k_1\in\{10,11,12\}$). 

Double remnants (\F~\ref{fig:st_col_double_out}) in our simulations are the result of CE evolution. After the CE, either a stripped He star ($k_1 \in \{7,8,9\}$) or a WD remains, with either a MS or WD companion.

\begin{figure}
\iftoggle{ApJFigs}{
\includegraphics[width=1.1\linewidth,trim = 10mm 10mm 0mm 0mm]{TEDI_m1s_CDF}
}{
\includegraphics[width=1.1\linewidth,trim = 10mm 10mm 0mm 0mm]{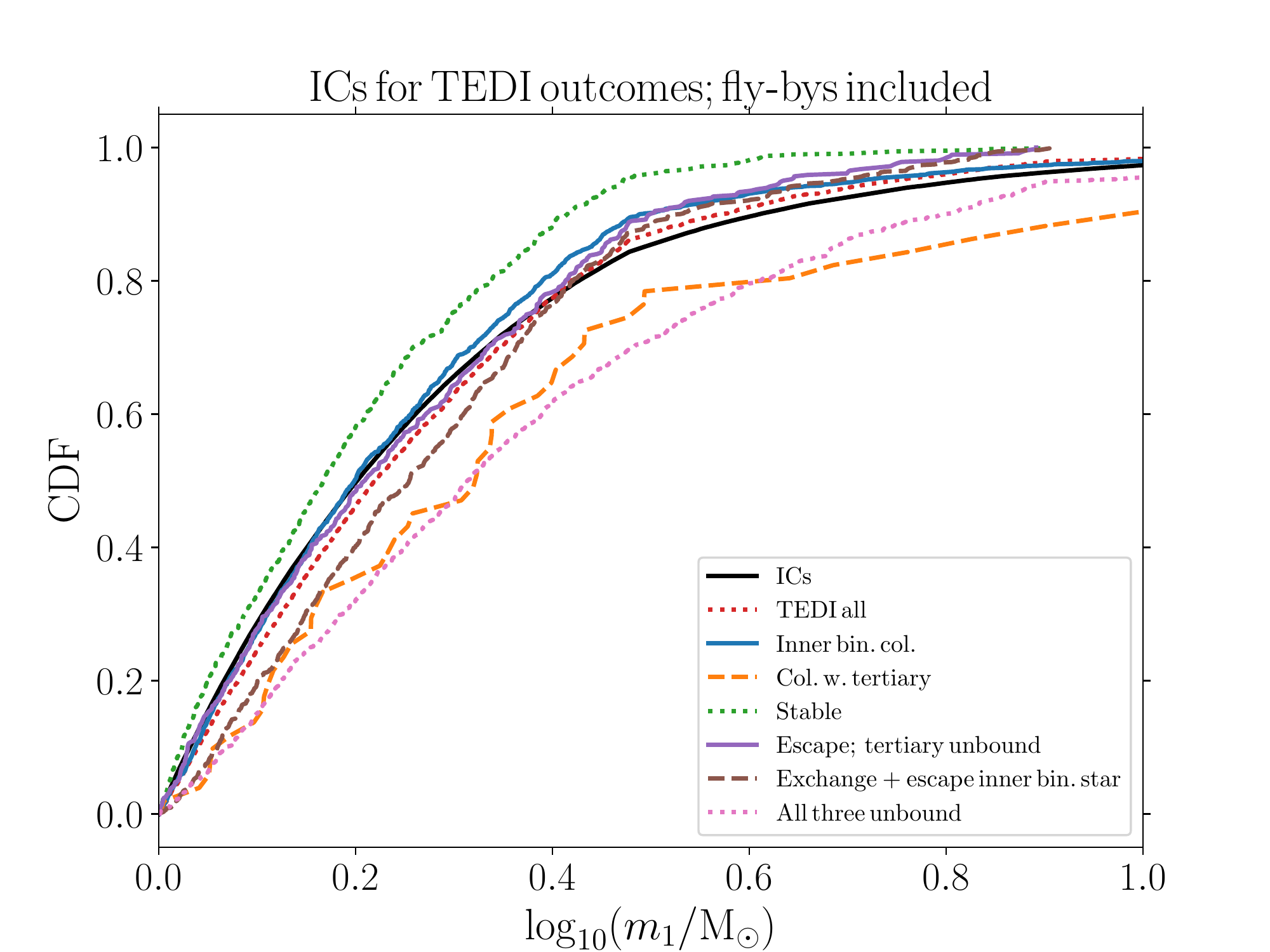}
}
\caption{Cumulative distributions of the initial inner binary primary mass ($m_1$) for different TEDI channels (colored lines). Also shown are the initial distribution with the solid black line (power law with $\mathrm{d}N/\mathrm{d} m_1 \propto m_1^{-2.35}$ for $m_1>1\,\msun$, cf. \S\ref{sect:popmeth:MC}). Refer to the caption for the meaning of the different colored lines. Fly-bys were included.}
\label{fig:m1s}
\end{figure}

\begin{figure}
\iftoggle{ApJFigs}{
\includegraphics[width=1.1\linewidth,trim = 10mm 10mm 0mm 0mm]{TEDI_q1s}
}{
\includegraphics[width=1.1\linewidth,trim = 10mm 10mm 0mm 0mm]{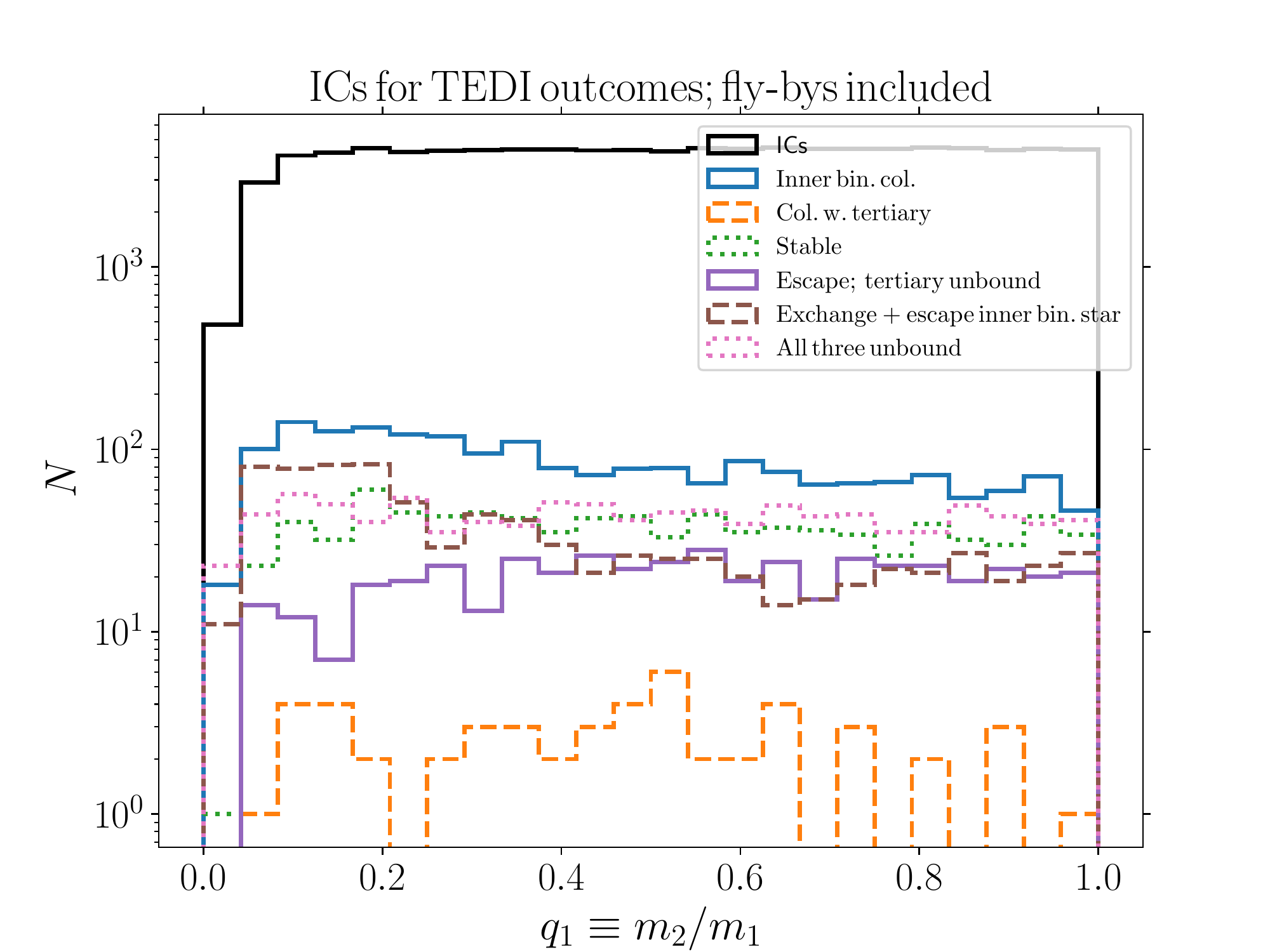}
}
\caption{Histograms of the initial inner binary mass ratio $q_1 \equiv m_2/m_1$ for different TEDI channels (colored lines). Also shown are the initial distribution with the solid black line, which is nearly flat (with an exception for low $q_1$, which arises because of the assumed lower limit on $m_2$). Refer to the caption for the meaning of the different colored lines. Fly-bys were included.}
\label{fig:q1s}
\end{figure}

\begin{figure}
\iftoggle{ApJFigs}{
\includegraphics[width=1.1\linewidth,trim = 10mm 10mm 0mm 0mm]{TEDI_q2s}
}{
\includegraphics[width=1.1\linewidth,trim = 10mm 10mm 0mm 0mm]{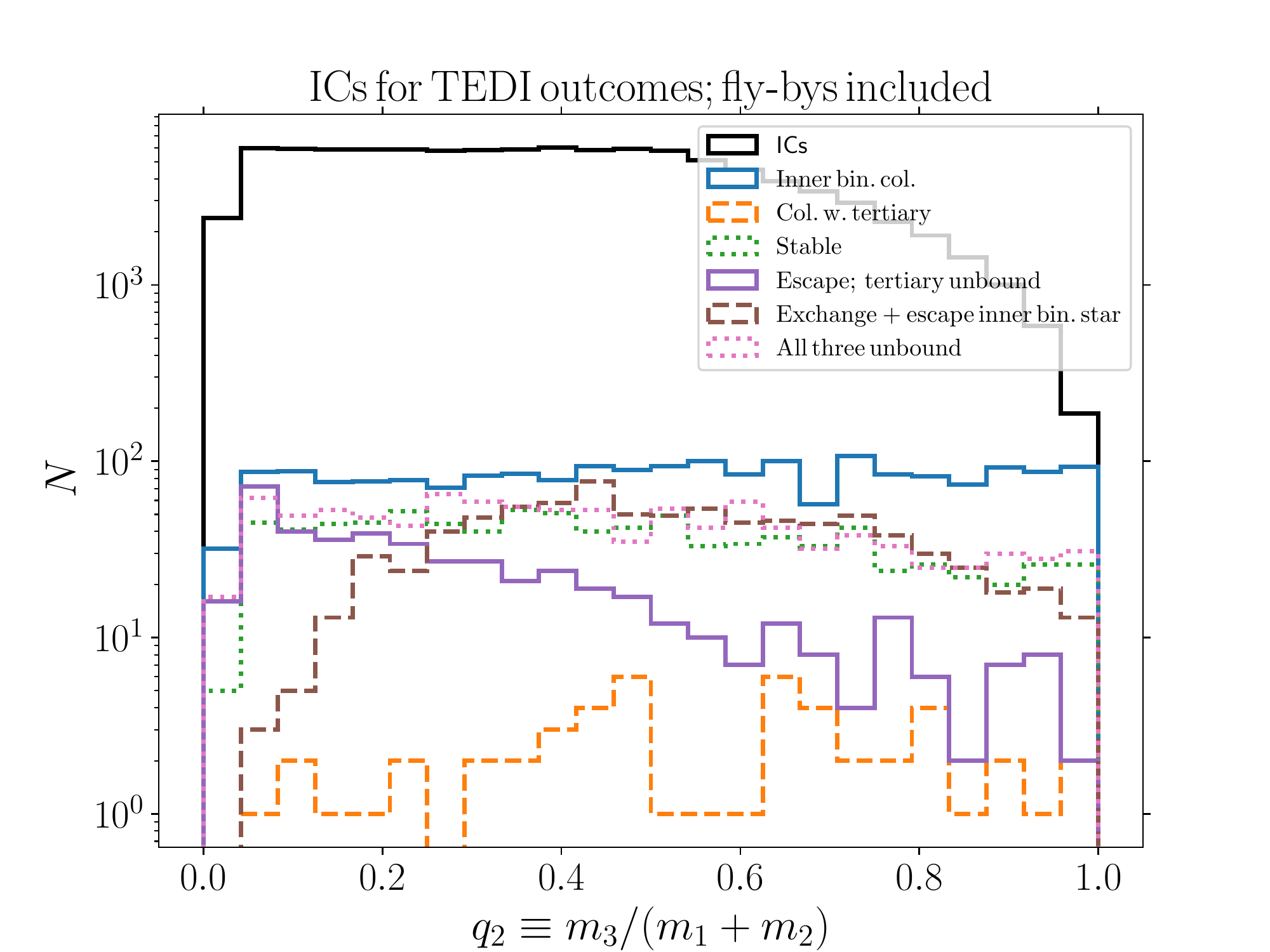}
}
\caption{Similar to \F~\ref{fig:q1s}, here showing the initial distribution of the outer mass ratio $q_2 \equiv m_3/(m_1+m_2)$. The initial distribution for all systems (solid black line) shows a paucity of systems with $q_2$ near unity, since the latter are more likely to be initially dynamically unstable (cf. \eq~\ref{eq:dynstab}). }
\label{fig:q2s}
\end{figure}

\subsection{Masses}
\label{sect:results:mass}
In the next few sections, we focus on the various outcomes of the TEDI according to the classification described in \S\ref{sect:popmeth:ch}. \F~\ref{fig:m1s} shows cumulative distributions of the initial inner binary primary mass $m_1$ for different TEDI channels. The initial distribution for all systems is shown with the solid black line (a power law with $\mathrm{d} N/\mathrm{d} m_1 \propto m_1^{-2.35}$ for $m_1>1\,\msun$, cf. \S\ref{sect:popmeth:MC}). Overall, all TEDI systems (red dotted lines in \F~\ref{fig:m1s}) closely follow the distribution of $m_1$ of all systems, indicating that the majority of TEDI systems are low-mass systems (since the assumed IMF implies that low-mass stars dominate). Notable exceptions are collisions involving the tertiary star which tend to favor somewhat higher $m_1$ (though we remark that collisions involving the tertiary star are exceedingly rare, cf. Table~\ref{table:rates}), and all three stars becoming unbound, which more strongly favors more massive primaries (a similar effect is seen in dedicated scattering experiments, see, e.g., \citealt{2016MNRAS.456.4219A}). 

In \F~\ref{fig:q1s}, we show histograms of the initial inner binary mass ratio $q_1 \equiv m_2/m_1$ for the different TEDI outcomes. We remark that we assumed a flat distribution in $q_1$, which is cut off at low $q_1$ ($q_1 \lesssim 0.1$) as a consequence of the assumed lower limit on $m_2$. Most TEDI outcome channels show similar distributions in $q_1$ compared to all systems (i.e., flat with a cut-off at low $q_1$). Notable exceptions are collision systems, which slightly favor lower $q_1$, and exchange with escape systems, which favor low $q_1$ more strongly. In the latter case, the preference for low $q_1$ can be attributed to the larger probability for the secondary star ($m_2$) to be ejected during an exchange interaction if it is of relatively low mass. 

\F~\ref{fig:q2s} shows similar histograms of the initial tertiary mass ratio $q_2 \equiv m_3/(m_1+m_2)$. Note that the initial distribution for all systems (solid black line) shows a paucity of systems with $q_2$ near unity, since the latter are more likely to be initially dynamically unstable (cf. \eq~\ref{eq:dynstab}). Small $q_2$ are clearly favored for the tertiary unbound channel, which is intuitively clear: relatively low mass tertiaries (compared to the inner binary) are easily ejected after instability. The exchange with escape channel (in which one of the inner binary stars escapes) shows a lack of systems with small $q_2$, since a relatively low-mass tertiary is more likely to be ejected from the system following instability, rather than form a new stable binary. 

\begin{figure}
\iftoggle{ApJFigs}{
\includegraphics[width=1.1\linewidth,trim = 10mm 10mm 0mm 0mm]{TEDI_col_mi_mf_CDF}
}{
\includegraphics[width=1.1\linewidth,trim = 10mm 10mm 0mm 0mm]{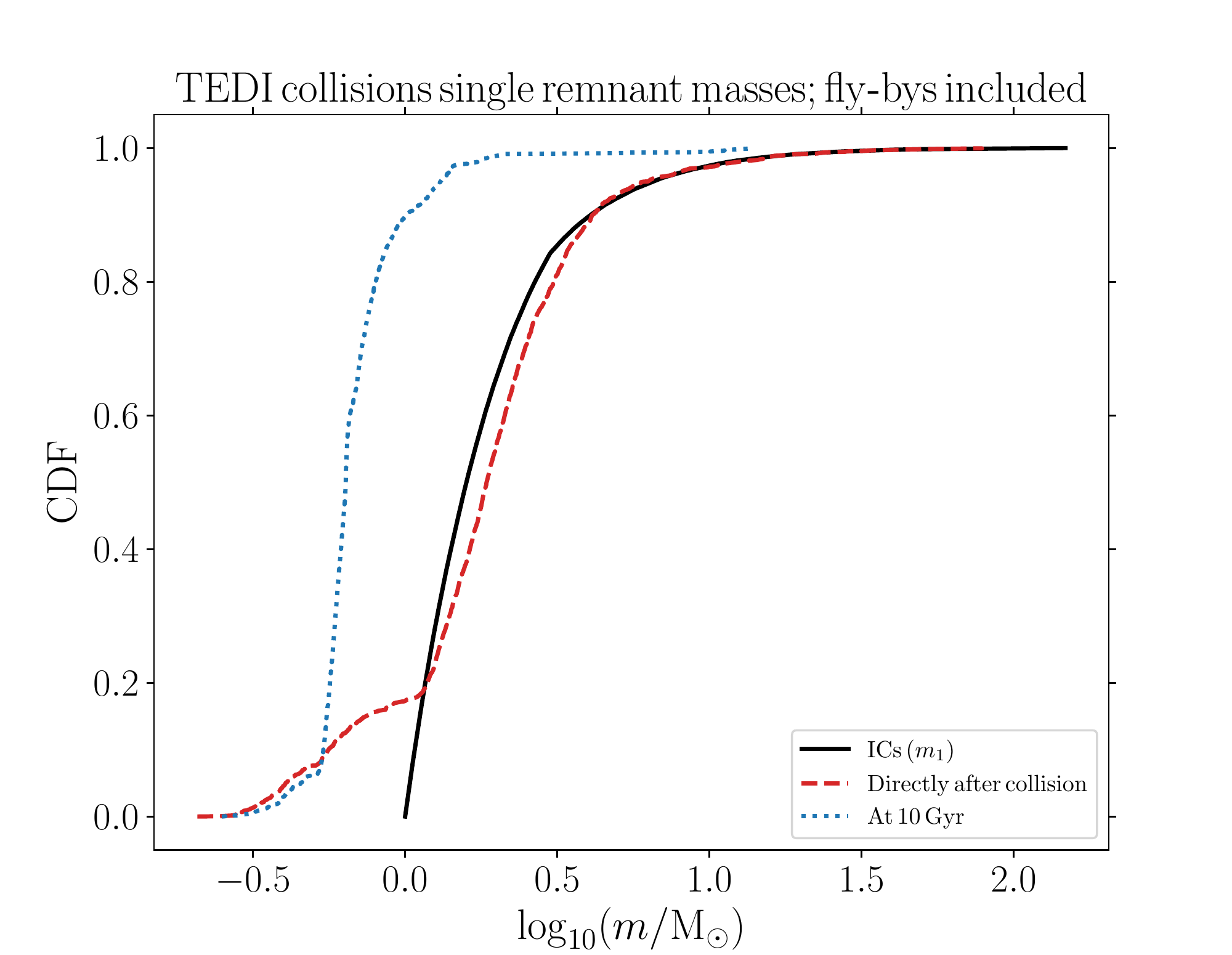}
}
\caption{Cumulative distributions of the masses for TEDI-induced collisions that resulted in a single object. Distributions are shown directly after collision (red dashed line), as well as after 10 Gyr of evolution (blue dotted line).  Also shown are the initial distribution with the solid black line (power law with $\mathrm{d} N/\mathrm{d} m_1 \propto m_1^{-2.35}$ for $m_1>1\,\msun$, cf. \S\ref{sect:popmeth:MC}). Fly-bys were included.}
\label{fig:m_col}
\end{figure}

In \F~\ref{fig:m_col}, we show cumulative distributions of the masses for TEDI-induced collisions that resulted into a single object. Distributions are shown directly after collision, as well as after 10 Gyr of evolution. Note that collisions will occur after a certain delay time, so colliding primary stars will have different masses by the time of collision due to stellar winds, mass transfer, and/or wind accretion. The median remnant mass directly after collision is only slightly higher than the median initial $m_1$; \F~\ref{fig:m_col} suggests that TEDI is not efficient at producing very massive stars. A caveat to this, however, is that high-mass stars are not well sampled in our simulations because of our choice of the initial primary mass range and the limited number of systems we are able to run given the computational expense.

\begin{figure}
\iftoggle{ApJFigs}{
\includegraphics[width=1.1\linewidth,trim = 10mm 10mm 0mm 0mm]{TEDI_orbits}
}{
\includegraphics[width=1.1\linewidth,trim = 10mm 10mm 0mm 0mm]{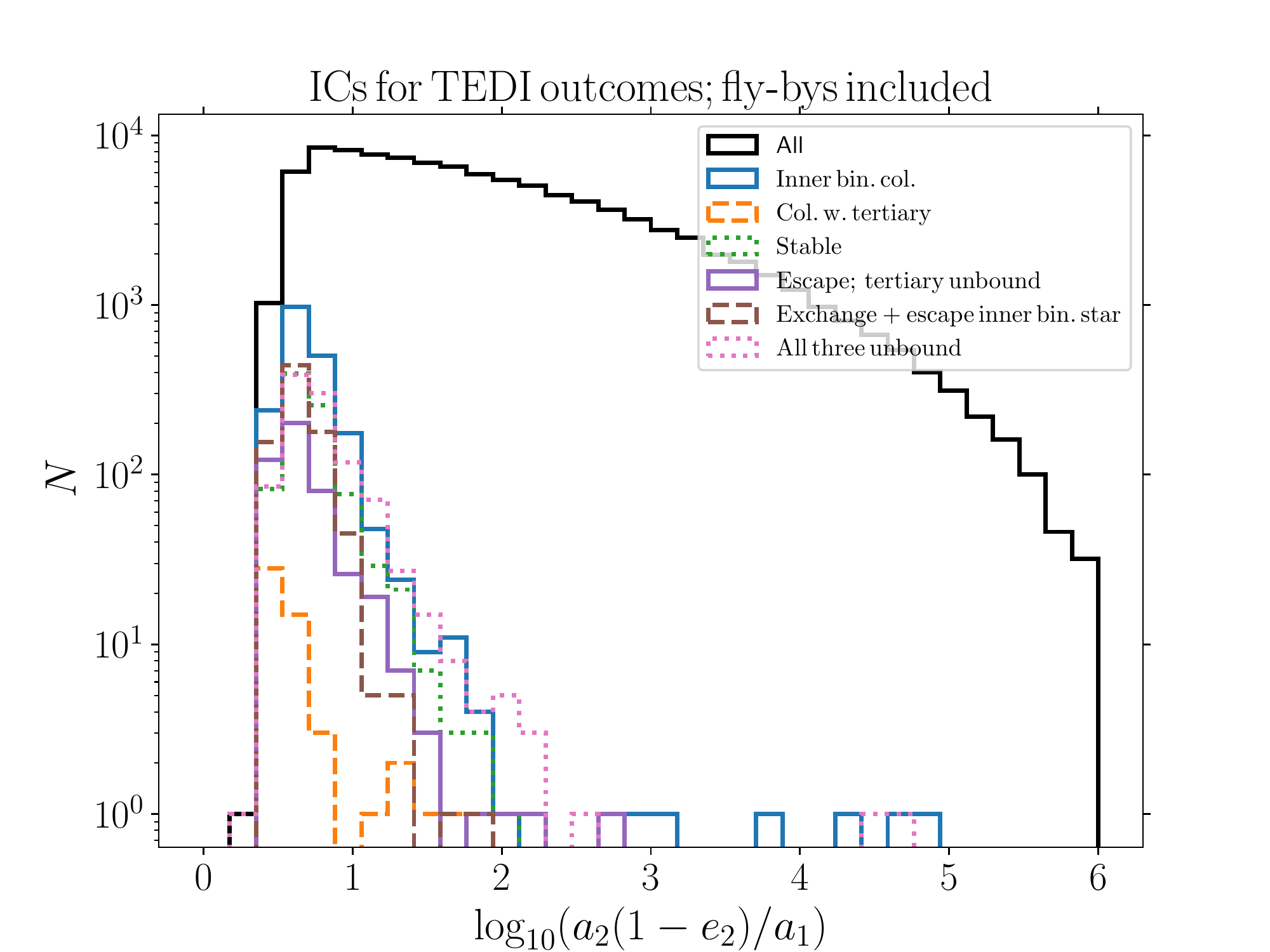}
}
\caption{Histograms of the initial ratio $a_2(1-e_2)/a_1$ for different TEDI systems (refer to the legend). The solid black line shows the initial distribution for all systems. Fly-bys were included. }
\label{fig:orbits}
\end{figure}

\subsection{Orbits}
\label{sect:results:orb}
For the TEDI outcomes, we show in \F~\ref{fig:orbits} the initial distribution of the `hierarchy parameter' $a_2(1-e_2)/a_1$. The majority of TEDI systems have small $a_2(1-e_2)/a_1$ peaking at a value of $\sim 5$, since they need to be sufficiently close to the boundary of instability in order to become unstable. Some systems show much larger values of $a_2(1-e_2)/a_1$. In these systems, instability is typically triggered by significant RLOF-induced inner orbital expansion, and/or changes in the outer orbit due to fly-bys.

\begin{figure}
\iftoggle{ApJFigs}{
\includegraphics[width=1.1\linewidth,trim = 10mm 10mm 0mm 0mm]{TEDI_times}
}{
\includegraphics[width=1.1\linewidth,trim = 10mm 10mm 0mm 0mm]{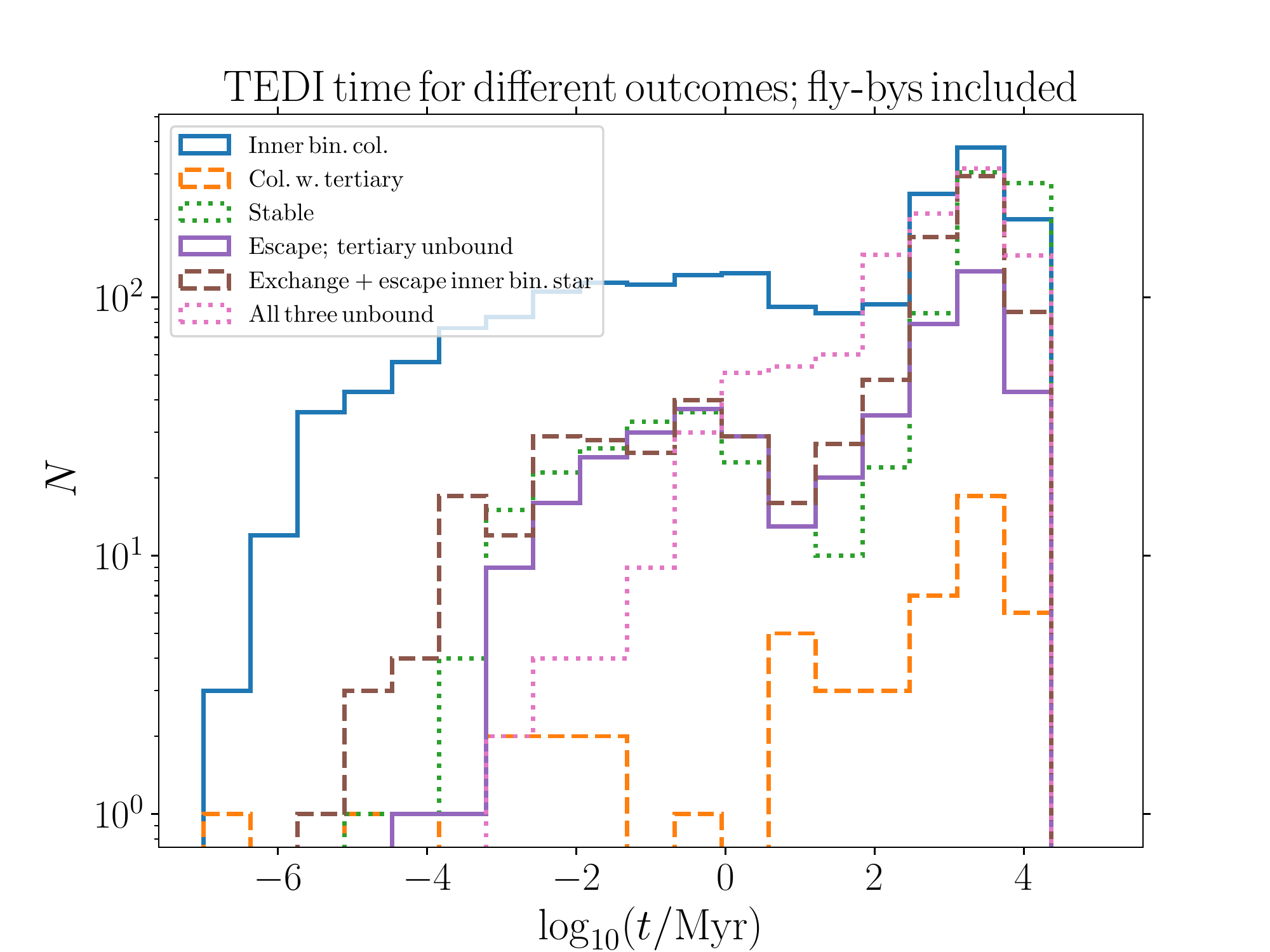}
}
\caption{Histograms of the time when TEDI was triggered, making a distinction between different outcomes (refer to the legend). Fly-bys were included.}
\label{fig:t}
\end{figure}

\subsection{Instability times}
\label{sect:results:t}
\F~\ref{fig:t} shows histograms of the times when instability was triggered in TEDI systems. Instabilities resulting in collisions can occur very early in the evolution (involving MS stars), and are due to systems that were initially only marginally stable (it could be argued that these systems are not `true' triple MS systems). However, for a significant fraction of collision systems the instability also occurs at much later times, up to $\sim 10\,\gyr$. With the exception of  MS-MS collisions, there are no notable differences in the instabilities times with respect to the various outcomes, which can be understood by noting that the underlying trigger for these outcomes (instability following stellar evolution, i.e., TEDI) is the same. 

The TEDI, producing collisions or ejections, will reduce the multiplicity fraction of triples with time, with potential implications for the observed time-dependent multiplicity fraction. A complication is that the TEDI is not the only pathway for reducing multiplicity in triples. For example, multiplicity is evidently also reduced when the inner binary merges (with or without aid from secular eccentricity excitation, but not due to the TEDI). With this caveat in mind, the peak of the TEDI delay time distributions near 1 Gyr implies that the multiplicity fraction should show a significant decrease for systems older than $\sim1\,\gyr$. A more detailed investigation that also takes into account higher-order systems is left for future work.

\begin{figure}
\iftoggle{ApJFigs}{
\includegraphics[width=1.1\linewidth,trim = 10mm 10mm 0mm 0mm]{TEDI_v_esc}
}{
\includegraphics[width=1.1\linewidth,trim = 10mm 10mm 0mm 0mm]{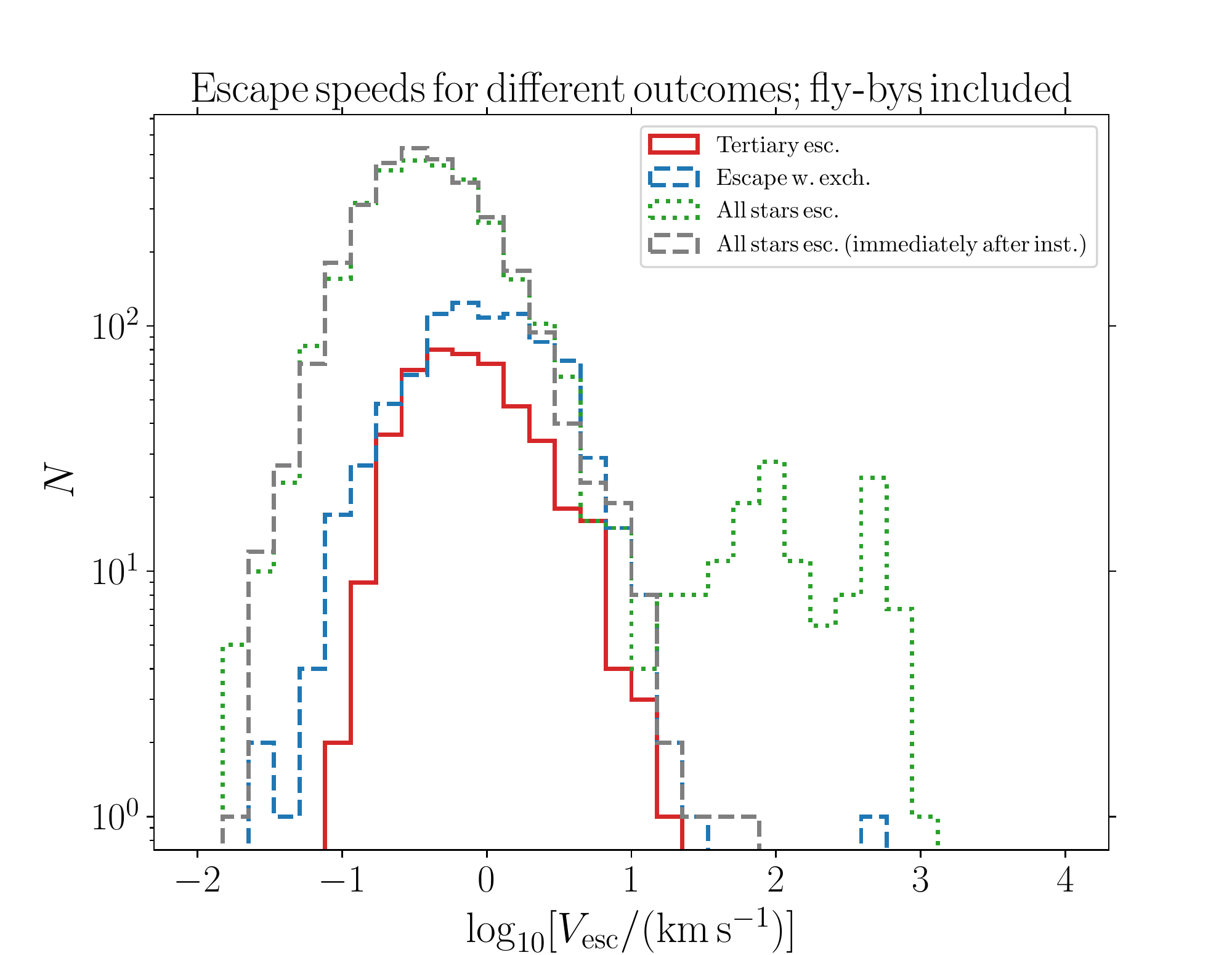}
}
\caption{Histograms of the escape speeds for three relevant TEDI channels: escape of the tertiary, escape of an inner binary star with the other inner binary star forming a stable binary with the tertiary, and all three stars becoming unbound (in the latter case, all three speeds are included). Escape speeds are recorded at 10 Gyr; for the channel of all three stars becoming unbound, we also show the distributions as recorded directly after instability (gray dashed line), showing that the high-speed tail is the result of SNe kicks. Escape speeds are measured relative to the initial centre of mass frame of the triple. Fly-bys were included. }
\label{fig:vesc}
\end{figure}

\subsection{Escape speeds}
\label{sect:results:vesc}
Lastly, we show in \F~\ref{fig:vesc} the distributions of escape speeds of unbound stars following TEDI. Except for the gray dashed line, the escape speeds were extracted at the end of the simulation (at a system age of $10\,\gyr$), implying that any potential effects of SNe kicks after the system become unbound are taken into account in the escape speeds, as well as --- and to a lesser degree --- changes in the velocity due to stellar winds (see \citealt{2021MNRAS.502.4479H}, Section 2.3). We distinguish between escape of the tertiary, escape of an inner binary star with the other inner binary star forming a stable binary with the tertiary, and all three stars becoming unbound (in the latter case, all three speeds are included; we also include the speeds as measured directly after instability, shown with the gray dashed line). Escape speeds are measured relative to the initial centre of mass frame of the triple. 

The distributions of the escape speed are similar for the two cases when only one star escapes (the tertiary star or an inner binary star); both are peaked around $1\,\kms$. In the case when all three stars become unbound, the distribution is wider, with both lower and higher escape speeds, and a lower peak value at $\sim 0.3\,\kms$ instead of $\sim1\,\kms$. When all three stars escape, this evidently includes the primary star, and the likelihood of any of the three stars receiving a significant SNe kick is higher, resulting in a higher speed tail extending up to $\sim 10^3\,\kms$. The latter tail is mainly shaped by SNe kicks, rather than interactions during dynamical instability, as is revealed by comparing the green dotted line in \F~\ref{fig:vesc} (which applies to all three stars becoming unbound, and recorded at 10 Gyr) to the gray dashed line (similar, but directly after instability). The high-speed tail consists of two local peaks near $\sim 100\,\kms$ (corresponding to BHs), and near $\sim 400\,\kms$ (corresponding to NSs).

Even when SNe kicks are included, hypervelocity stars, i.e., stars with escape speeds $>10^3\,\kms$, are not produced, but this conclusion is evidently dependent on our assumption of the SNe kick distribution (here, we assumed a Maxwellian distribution with dispersion $\sigma_{\mathrm{kick}} = 265\,\kms$ for NSs and $\sigma_{\mathrm{kick}} = 50\,\kms$ for BHs, see \S\ref{sect:meth:stel}). Although not shown in \F~\ref{fig:vesc}, we remark that no escaping MS stars in our simulations have escape speeds in excess of $\sim 30 \, \kms$.

\section{Discussion}
\label{sect:dis}

\subsection{MS mergers}
\label{sect:dis:ms}
In contrast to \pk, we find that the majority of collisions following TEDI involve two MS stars. This can be attributed to the fact that \pk~did not take into account the coupled evolution of the inner binary with the tertiary star. Specifically, an initially only marginally stable triple can become unstable quickly, leading to early collisions involving MS stars. Similarly, strong secular evolution during the MS can transition the system to the semisecular regime \citep[e.g.,][]{2012ApJ...757...27A,2014ApJ...781...45A,2016MNRAS.458.3060L,2018MNRAS.481.4907G,2018MNRAS.481.4602L,2019MNRAS.490.4756L,2020MNRAS.494.5492H} which we here consider as part of the TEDI, and which can also lead to early collisions. Furthermore, instability can be triggered by mass transfer during the MS, when the inner binary donor has become less massive than the donor, in particular if the orbit is eccentric. Despite these differences between our work and \pk, our collision rates for evolved stars, in particular involving giant stars, are $\sim 1\times  10^{-4}\,\yr^{-1}$ and are therefore consistent with the overall rate of $1.2\times 10^{-4}\,\yr^{-1}$ found by \pk. This suggests that our MS-MS collision rates are {\it in addition} to those found by \pk~for more evolved stars.

Our high rates of MS collisions imply that dynamical instability in triple stars is efficient at producing blue straggler stars, which are MS stars that appear to be too blue and luminous compared to the MS turnoff point of the cluster that they reside in \citep[e.g.,][]{1953AJ.....58...61S,1993PASP..105.1081S,1995ARA&A..33..133B,1997ApJ...487..290S,1999ApJ...513..428S,2001ApJ...548..323S,2002MNRAS.332...49S,2013ApJ...777..106C}. Blue stragglers could be stellar merger products, in particular as a result of secular evolution in triples \citep{2009ApJ...697.1048P}; the mergers found in this work are strictly speaking a different channel, since they involve dynamical instability. Simulations indicate that the formation rate of blue stragglers as a result of chance collisions in globular clusters is on the order of a few per $\gyr$ \citep[e.g.,][]{2019A&A...621L..10P} whereas our TEDI-induced MS-MS merger rate is $\sim 10^{-4}\,\yr^{-1}$, suggesting that TEDI-induced blue straggler formation in the Galaxy could be very efficient. Moreover, blue straggler stars produced through the TEDI could still have wide companions with apparently different age. A more careful and quantitative comparison to observations should be considered in future work. 

Another implication of a high rate of MS-MS collisions might be the formation of massive MS stars which could ultimately produce overmassive BHs. The latter is particularly interesting in light of recent detections of GWs from high-mass BH mergers, most notably the GW source GW190521 with a remnant BH mass of $142_{-16}^{+28} \, \msun$ \citep{2020PhRvL.125j1102A}. In our simulations, however, we do not find that TEDI produces a large population of highly massive stars (see \F~\ref{fig:m_col}). A caveat to this is that the majority of systems in our simulations were of lower mass as a result of the IMF and choosing a relatively low cut-off mass for the primary star, $m_1>1\,\msun$. 

Generally, TEDI-induced collisions could lead to wide binaries with apparently asynchronous ages, or other peculiarities. This includes wide binaries with He WDs (cf. \F~\ref{fig:st_col_single_out}) which would otherwise be difficult to explain through standard isolated binary evolution.

\subsection{Eccentric CE}
\label{sect:dis:ece}
In our simulations, the majority of collisions involve giant stars with extended (convective) envelopes, and are therefore expected to likely lead to CE evolution (cf. Table~\ref{table:gen_fractions}). However, given the high relative velocity and small impact parameter, the outcome of CE evolution in these cases could be quite different; in particular, tight binaries could remain with significant residual eccentricity \citep[e.g.,][]{2021arXiv210502227G}, producing eccentric post-CE binaries with evolved/WD components such as the Sirius WD-MS binary (e.g., \citealt{2005ApJ...630L..69L,2008A&A...480..797B}; \pk; \citealt{2017ApJ...840...70B}).

\subsection{WD collisions}
\label{sect:dis:wdc}
TEDI-induced collisions in our simulations include colliding WDs, which could potentially produce SNe Ia \citep[e.g.,][]{2009ApJ...705L.128R,2010MNRAS.406.2749L,2011A&A...528A.117P,2016ApJ...822...19P}. WD collisions could also occur as a result of (strong) secular evolution \citep[e.g.,][]{2011ApJ...741...82T,2012arXiv1211.4584K,2013MNRAS.430.2262H,2013ApJ...778L..37K,2018A&A...610A..22T,2018MNRAS.478..620H,2019ApJ...882...24H}. 

Our estimated Galactic TEDI WD collision rates are $\sim 10^{-6}\,\yr^{-1}$ (cf. Table~\ref{table:rates}). In comparison, the observed Galactic SNe Ia rate is $\sim 10^{-2}\,\yr^{-1}$ \citep{2013ApJ...778..164A}, indicating that TEDI-induced WD collisions likely do not contribute significantly to Galactic SNe Ia.

\subsection{NS and BH mergers}
\label{sect:dis:ms}
We did not find TEDI-induced collisions involving NSs and/or BHs in our simulations (cf. \F~\ref{fig:st_col}, top panel). Progenitor systems of NSs and BHs suffer from strong mass loss and natal kicks, which tend to disrupt the system quickly, rather than triggering a dynamical instability phase (with lower relative velocities). However, we do not completely exclude the possibility of TEDI-induced NS and BH mergers, since we focussed on lower-mass stars. Better statistics in the high-mass range could be attained by further restricting the mass range to high-mass stars, e.g., $m_1>8\,\msun$, but this is beyond the scope of this work.

\subsection{Future directions}
\label{sect:dis:fut}
In addition to investigating the high mass-end range in more detail by further restricting the initial primary mass range, we mention here a number of other potential future directions. These include investigating different distributions for natal kicks, which will be of importance for NSs and BHs. Also, different metallicities will change the details of wind mass loss and the stellar radii evolution, which could in turn be important for the TEDI scenario. Lastly, as is clear from our results (\S\ref{sect:results:rate:res}), the TEDI, although interesting, has only a small contribution to all mergers in stellar triples; the overall population of mergers (cf. \F~\ref{fig:st_col}, bottom panel) should be considered in more detail in future work.

\section{Conclusions}
\label{sect:conclusions}
Using a state-of-the-art population synthesis code including stellar evolution, binary interactions, and gravitational dynamics, we have revisited the Triple Evolution Dynamical Instability (TEDI) channel. In the TEDI, mass loss in evolving triples triggers short-term dynamical instabilities, which can lead to head-on collisions of stars, exchanges, and ejections. Our main conclusions are listed below. 

\medskip \noindent 1. Based on population synthesis calculations with \mse~\citep{2021MNRAS.502.4479H}, our estimated stellar collision rate in the TEDI channel (taking into account fly-bys in the field) is $\simeq 2.5 \times 10^{-4}\,\yr^{-1}$ with MS-MS collisions included, and $\simeq 0.9 \times 10^{-4} \, \yr$ for collisions involving more evolved stars, consistent with the previous work of \citet{2012ApJ...760...99P}, whose simulations included fewer effects. Our rate is dominated by collisions of MS stars, although collisions with more evolved stars, including giants and WDs, also occur in significant numbers.

\medskip \noindent 2. When a TEDI-induced collision results in a single merger remnant, the most likely outcome is another MS star, implying that the TEDI is potentially efficient at producing blue straggler stars (resulting from head-on collisions). Other likely single outcomes include single giant stars, single WDs, and potentially SNe Ia when two WDs collide head-on. When a CE is triggered following collisions in our simulations, the outcome of the collision can be two stars in a close (and potentially still eccentric) orbit; in the latter case, the most likely outcomes are stripped He stars with MS stars, WDs with MS stars, and double WD systems. Specifically, the latter could explain eccentric post-CE systems.

\medskip \noindent 3. In addition to collision rates, we estimated rates of other outcomes of the TEDI including the tertiary star becoming unbound, an exchange interaction during which on of the inner binary stars escapes and the other inner binary star forms a new binary with the tertiary star, and all three stars becoming unbound. These outcomes have Galactic event rates of $\sim 0.6$, $\sim 1$, and $\sim 1.3 \times 10^{-4}\,\yr^{-1}$ for the three channels, respectively.

\medskip \noindent 4. When fly-bys are included, all TEDI rates increase systematically; for collisions, the increase is $\sim 17\%$. Fly-bys can decrease the outer orbital periapsis distance, increasing the likelihood for dynamical instability in the system by $\approx 31\%$ in our simulations (excluding semisecular cases). 

\medskip \noindent 5. As expected based on gravitational dynamics arguments, the TEDI outcome with the tertiary star becoming unbound favors systems with initially low tertiary mass ratio $q_3 =m_3/(m_1+m_2)$. On the other hand, exchange interactions with an inner binary star escaping and the tertiary forming a new binary system favor low initial inner binary mass ratio $q_1 = m_2/m_1$. 

\medskip \noindent 6. Escaping stars following the TEDI in our simulations have typically low escape speed (with respect to the initial centre of mass frame of the triple). When the tertiary escapes or an exchange interaction occurs, the typical escape speed is $\sim 1\,\kms$. When all three stars escape, the escape speed tends to be slightly lower on average, with a peak near $\sim 0.3 \, \kms$, although there is also a high-speed tail associated with SNe kicks, up to $\sim 10^3\,\kms$ (and depending on our assumed SNe kick prescription). However, if SNe kicks are not considered, then the TEDI is unable to produce stars with escape speeds above $\sim 100\,\kms$, i.e., we do not expect the TEDI to be an important mechanism for producing hypervelocity stars.

\begin{acknowledgements} 
We thank the anonymous referee for a helpful report. A.S.H. thanks the Max Planck Society for support through a Max Planck Research Group.
\end{acknowledgements}

\bibliographystyle{aasjournal}
\bibliography{literature.bib}


\end{document}